\documentclass{book}

\usepackage{fancyhdr}
\fancypagestyle{plain}{%
\fancyhead{} 
}

\usepackage{setspace} 
\usepackage[english]{babel}
\usepackage{amsfonts,amsthm}
\usepackage{amsmath}
\usepackage{amssymb}
\usepackage[all]{xy}

\usepackage{epsfig}

\theoremstyle{definition} 
\newtheorem{definition}{Definition}
\newtheorem{postulate}{Postulate}

\theoremstyle{plain} 
\newtheorem{theorem}{Theorem}
\newtheorem{lemma}[theorem]{Lemma}
\newtheorem{corollary}[theorem]{Corollary}

\theoremstyle{remark}
\newtheorem{example}{Example}[section]

\newcommand{\ket}[1]{|#1\rangle}
\newcommand{\bra}[1]{\langle#1|}
\newcommand{\braket}[2]{\langle#1|#2\rangle}
\newcommand{\expec}[1]{\langle#1\rangle} 

\newcommand{\C}[0]{\mathbb{C}^d}
\newcommand{\B}[0]{{\cal B(H)}} 
\newcommand{\Bsa}[0]{{\cal B(H)_\mathrm{sa}}} 
\newcommand{\Hi}[0]{{\cal H}} 
\newcommand{\PH}[0]{{{\cal P}_{d-1}}} 
\newcommand{\PO}[0]{{{\cal P}_1}} 
\newcommand{\D}[0]{\mathrm{d}}

\makeatletter
\def\thickhrulefill{\leavevmode \leaders \hrule height 1pt\hfill \kern \z@}
\newcommand{\tp}{\begin{titlepage}%
    \let\footnotesize\small
    \let\footnoterule\relax
    \parindent \z@
    \reset@font
    \null\vfil
    \hrule height 4pt
    \vskip 10\p@
    \begin{flushright}
      \LARGE 
      \strut \@title \par
      \vskip 30\p@
      \strut \@author
    \end{flushright}
    \vskip 5\p@
    \hrule height 4pt
    \vskip 60\p@
    \vfil\null
  \end{titlepage}%
  \setcounter{footnote}{0}%
}

\makeatother

\title{\Huge{Optimal Trade-Off}\\
			 \Large{Information Gain vs Distortion Loss\\
			 In Finite-Dimensional Quantum Systems}}
\author{Thijs van der Valk}
\date{December 2005}

\begin{document}
	\frontmatter
\tp

\cleardoublepage

\vspace{3cm}

\begin{center}
{\Large Master's thesis by Thijs van der Valk}\\
	Department of Mathematical Physics\\
	Radboud University Nijmegen\\
	Supervisor: Hans Maassen\\
\vspace{1.5cm}
	Nijmegen, December 2005
\end{center}

\vspace{6cm}

\begin{figure}[!hb]
\begin{center}
\includegraphics[]{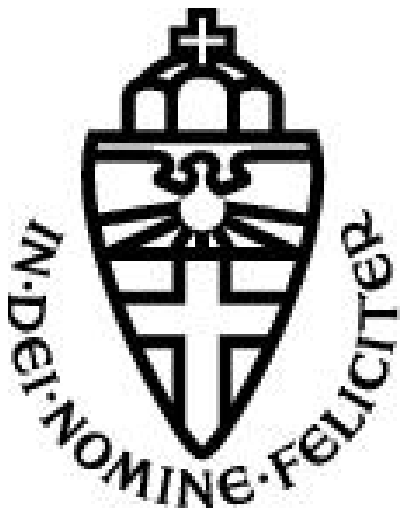}
\end{center}
\end{figure}

\thispagestyle{empty}

\cleardoublepage

\vspace{12cm}

\begin{center}
De bink is binnen. (Jan Cremer, ``Ik, Jan Cremer'', 1961)
\end{center}

\begin{center}
Do you trust me? (Jack Bauer, ``24'', 2004)
\end{center}

\thispagestyle{empty}

\cleardoublepage

\chapter{Introduction}
\section{Heisenberg's Uncertainty Relations}

In 1927, Werner Heisenberg (1901-1976), founding father of quantum mechanics, wrote a paper called \emph{\"Uber den anschaulichen Inhalt der quantentheoretischen Kinematik und Mechanik}~\cite{Hei}. In this paper he introduced the famous uncertainty relations, nowadays referred to as the Heisenberg uncertainty relations. The relations express the impossibility to measure certain pairs of observable variables at the same time, with infinite precision. The most cited and appealing one is
\begin{equation}
\Delta p \Delta q \geq \frac\hbar 2,
\end{equation}
in which $\Delta p$ is the variance of the momentum and $\Delta q$ the variance of the position of some particle. In the same paper, he derived similar results for time and energy, and action and angle. Although Heisenberg himself never referred to the relations as a principle explicitly, the term uncertainty principle came into vogue shortly after publication.

\subsubsection{An Analogy}

An anthropologist, named Esther, desires to do research on the behavior of some primitive, pre-modern society. Esther, as she wants to find out how natives think and act, has to participate in every day life. This is of course hard; participation implies distortion, since Esther's presence will undoubtedly influence the behavior of the natives. Quantum measurement is like this socio-cultural measurement: there exists interaction between observation and distortion.

The difference between the distortion as a consequence of Esther's research and the distortion imposed by measurement of microscopic quantum particles, is the fundamental nature of the latter. 

\begin{figure}[!htb]
\begin{center}
\includegraphics[width=8cm]{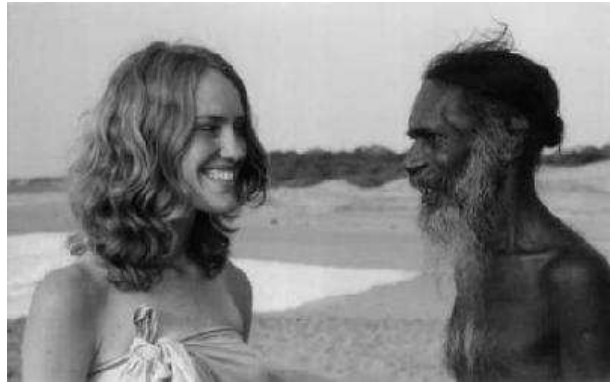}
\caption{Anthropologist in action.}
\label{fig:antro}
\end{center}
\end{figure}

\begin{figure}[!htb]
\begin{center}
\includegraphics[width=4cm]{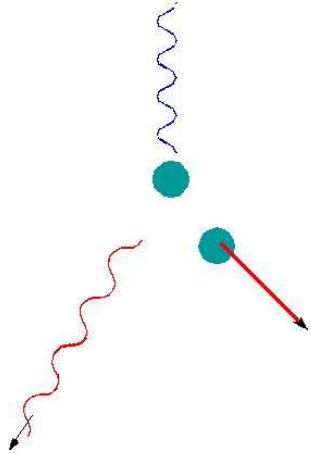}
\caption{The Compton Effect is due to the collision interaction between a photon and an electron. Light from the microscope hits an electron. Both photon and electron are scattered and obtain new momenta.}
\label{fig:compt}
\end{center}
\end{figure}

In his 1927 paper, Heisenberg considers the measurement of the position of an electron by a microscope. Since electrons are so small, they can only be observed with use of high-energy or short-wavelength light, e.g. a X-rays microscope is needed. At high-energy scales however, the Compton effect cannot be ignored (see fig.~(\ref{fig:compt})). 

The collision with light particles changes the momentum of the electron. So measurement of the position, results in distortion of the momentum. 

\begin{quote}
At the instant of time when the position is determined, that is, at the instant when the photon is scattered by the electron, the electron undergoes a discontinuous change in momentum. This change is the greater the smaller the wavelength of the light employed, i.e., the more exact the determination of the position. At the instant at which the position of the electron is known, its momentum therefore can be known only up to magnitudes which correspond to that discontinuous change; thus, the more precisely the position is determined, the less precisely the momentum is known, and conversely. (Heisenberg, 1927) 
\end{quote}

Closer observation involves more distortion. The Heisenberg relations lay fundamental boundaries upon the amount of information that can be extracted by observation, and distortion due to the same observation. The distortion as a consequence of Esther's observation of the pre-modern society is not fundamental. It is a result of improper methods. Instead of physical participation in native life, she could use other research methods, such as (hidden) camera's or questionairies. Esther could argue, and she will, that observation of that kind doesn't suffice to understand the aboriginal people. To the extent that that is true, there are indeed limits to the pair of information gain and distortion loss in this kind of research. These limits are not of fundamental, but of practical nature. There might exist anthropological observation methods that circumvent distortion. Moreover, as opposed to a pure quantum system, there will always be aspects that can be measured without distortion, even for very complex systems as a primitive society. I wonder if this in spite of or thanks to this complexity; measurement of the position of one electron seems to be harder than measurement of the hunting customs of aboriginals in which a few more of these little fellows are present (so I heard).

Nevertheless, the Heisenberg uncertainty principle plays an important role in modern science. It reflects on the position of natural sciences in our post-modern world.

\begin{quote}
Natural science, does not simply describe and explain nature; it is part of the interplay between nature and ourselves. (Heisenberg, Physics and Philosophy, 1963)
\end{quote}

\subsubsection{The Heisenberg Uncertainty Principle}
The Heisenberg principle for an arbitrary quantum system is faulty formulated as

\begin{quote}
It is impossible to extract information from a quantum system without changing its state.
\end{quote}

It is a faulty formulation, since if we realize that a state covers the expectation values of a quantum system, it is naturally that information extraction implies state change. For example consider tossing a die in dice cup. After shaking and before looking in the cup, the die is in a completely mixed state; any side of the die can be up. If you open the cup, it is clear which side is up and you changed the state from fully mixed to pure. 

A better formulation is:

\begin{quote}
There will always exist at least one state, such that, if the system is measured, i.e. information is extracted, and this information is disregarded, this state will be changed. 
\end{quote}

The Heisenberg principle does not assert that all states are changed. For example, consider a spin-1/2 particle with its spin in a certain $\vec z$-direction. Measurement of the spin in the $\vec z$-direction will not change the state. Furthermore, notice that a die in a dice-cup does not obey the Heisenberg principle; the state of a die will not change if you close the cup again after opening it and forgetting what you saw. 

\newpage

\section{This Thesis}
The pair consisting of information gain and distortion loss is restricted by the Heisenberg principle. The goal of my research project was to construct a general mathematical formulation of the Heisenberg principle. In other words, the goal was to answer in mathematical terms the question ``What is the maximal amount of information that can be extracted from a system if the amount of distortion is fixed?'' Or better, ``What is the trade-off between information gain and distortion loss?''

In the optimal trade-off between these two entities, there are two extremes: absolute containment of an initial system, so no information extraction, and maximal information extraction, so no containment. 

The former is of course easy to realize: leave the system untouched. The latter is harder and is what is called \emph{optimal state estimation}. This is optimal measurement followed by an optimal guess and leads to an explicit procedure to find the best estimation of the initial system. The loss of information about the initial system, the distortion, is unavoidable by the Heisenberg principle. It is controllable though as it depends crucially on the measuring procedure.

\begin{figure}[!htb]
\begin{center}
\includegraphics[width=4cm]{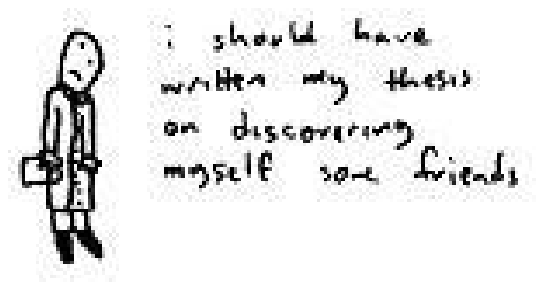}
\label{fig:mythesis}
\end{center}
\end{figure}

Finding this crucial dependence and so obtaining the physical restrictions to the pair of information gain and distortion loss, led to an explicit uncertainty relation. It led in particular to a class of optimal instruments that saturate this Heisenberg relation. In addition, the families of covariant quantum operations, covariant POVMs and covariant measurement instruments were classified. Examples of covariant devices are optimal spin-flip devices~\cite{Wer3,Gis}, optimal pure-state cloners~\cite{Wer1} and optimal estimation devices~\cite{Mas}. 

This thesis is the result of 9 months of research. It starts in chapter~\ref{chap:qm} with an introduction to quantum mechanics as I understand it. Furthermore, it contains some tools needed in the chapters thereafter. 

In chapter~\ref{chap:qmeas}, I classify the families of covariant quantum operations, POVMs and measurement instruments. It yields a one-parameter class of covariant quantum operations, a four-parameters class of covariant POVMs and a ($4+1$)-parameter class of covariant measurement instruments.

Finally chapter~\ref{chap:trade} contains the joint-optimization of the different classes of operations, leading to an uncertainty relation that restricts the pair of information gain and distortion loss of an arbitrary finite-dimensional quantum system. This chapter contains the main theorem of this thesis. 
\newpage

\section{Thanks}

I would to like to thank my mother Joke, my father Pieter, my sister Hanna, my two little brothers Joost and Dirk, my girlfriend Esther, my good friend Arnout (for letting me wake him up at noon), my friends, my supervisors Dr. Hans Maassen and Dr. M\u ad\u alin Gu\c t\u a, and second corrector Prof. Dr. Ronald Kleiss. 
\cleardoublepage

\renewcommand{\contentsname}{Table of Contents}  
\begin{spacing}{1.2}  
\tableofcontents%
\cleardoublepage 
\end{spacing} 

\mainmatter

\chapter{Quantum Mechanics}
\label{chap:qm}

This chapter starts with a brief introduction to quantum mechanics (section~\ref{sec:qm}) and quantum operations (section~\ref{sec:qo}). In section~\ref{sec:noclone} the no-cloning theorem will be treated. This theorem, as an example of a quantum operation, illustrates the interconnection between mathematics and physics; it shows how theoretical results of mathematical physical research are used in a physical framework. In the last section~\ref{sec:mt}, some mathematical tools that are needed in chapter~\ref{chap:qmeas} and chapter~\ref{chap:trade}, are elaborated. 
\section{Quantum Mechanics}
\label{sec:qm}
Quantum mechanics has been formulated in many different languages~\cite{Emch}. Most famous are Heisenberg's matrix mechanics and Schr\"o\-din\-ger's wave mechanics. These two together, which Schr\"o\-din\-ger in 1926 pointed out to be equivalent, were united and given a firm formal foundation in the Hilbert space formulation. It was established between 1926 and 1933 as an accumulation of several books and articles written by (nowadays) famous physicists and mathematicians.

\subsection{Von Neumann's Hilbert Space Formulation}
One of the most important books to appear in that era, is Johann von Neumann's \emph{Grundlagen der Quantenmechanik}~\footnote{According to N.P. Landsman the quantum mechanical equivalent of Newton's \emph{Principia}}. It provides us with an axiomatic approach to quantum mechanics. 

\begin{postulate}[Von Neumann's postulate]
A physical system is described by a triplet $\{S,{\cal A},<.;.>\}$, where: $S$ is the set of the possible states of the system; $\cal A$ is called its algebra of observables; and $<.;.> : S\times{\cal A}\mapsto \mathbb R$ is the prediction rule corresponding to the expectation value of the observable $A\in{\cal A}$ when the system is in state $\psi\in S$.
\end{postulate}

A quantum system is by assumption described by a separable Hilbert space $\Hi$. This means that it has a countable, orthonormal basis. The observables are contained in the algebra of bounded operators $\B$ on $\Hi$. The observables are the Hermitian elements of this algebra and in general do not form an algebra.  States are identified with the set of all positive, trace-class operators $\hat{\rho}$ on $\Hi$. They are called density operators and are normalized by $\mathrm{Tr}(\hat{\rho})=1$, where $\mathrm{Tr}$ is the trace. The prediction rule $<.;.>$ is defined by $\mathrm{Tr}(\hat{\rho} A)$. 

Famous in Hilbert space formulation is Paul Dirac's \emph{bra-ket}-notation. Functionals on a Hilbert space $\Hi$ are denoted by a \emph{bra} $\bra{\psi}$. Vectors of $\Hi$ are denoted by a ket $\ket{\phi}$. The bra-ket itself, $\braket{\psi}{\phi}$, is defined by the standard inner product on $\Hi$, $\braket{\psi}{\phi}\equiv(\ket{\psi},\ket{\phi})$. This notation is justified by the Riesz representation theorem, since this theorem states that every Hilbert space $\Hi$ is isometrically isomorphic to its dual space $\Hi^*$. The dual space is the space of all functionals of $\Hi$. Therefore there exists a unique $\phi\in\Hi$ for every $\psi\in\Hi^*$ such that $\psi{\theta}=({\phi},{\theta})$ for all $\theta\in\Hi$. The functional $\psi$ is denoted by the bra $\bra\phi$.

\subsubsection{Time-evolution}
The symmetries that express the dynamics of a quantum system are covered by one-parameter groups of Kadison automorphisms. These are defined as bijective maps $\alpha$ of the set $S$ onto itself, satisfying 
\begin{equation}
\alpha(\lambda \hat{\rho}_1 + (1-\lambda)\hat{\rho}_1)=\lambda\alpha(\hat{\rho}_1)+(1-\lambda)\alpha(\hat{\rho}_2).
\end{equation}
Unitarity of time-evolution is obtained by Wigner's theorem, which proves that every automorphism $\alpha$ is of the form 
\begin{equation}
\alpha(\hat{\rho})=U\hat{\rho} U^*
\end{equation}
with $U$ a unitary or anti-unitary map, uniquely determined up to a phase. The Schr\"odinger equation, 
\begin{equation}
i\hbar \frac{\D}{\D t} \ket\psi=H\ket\psi,
\end{equation}
is asserted by Stone's theorem.

\begin{theorem}[Stone's Theorem]
\label{the:stone}
Let $t\to U(t)$ be a strongly continuous map from $\mathbb{R}$ to the unitary operators so $U(t+s)=U(t)U(s)$. Then $U=e^{-iHt}$ for a unique Hermitian operator $H$. 
\end{theorem}

The operator $H$ in theorem~\ref{the:main} is called the Hamiltonian. See~\cite{Simon} for an interesting and readable treatise on quantum dynamics. 

\subsection{The $C^*$-algebraic Formulation}
Not all physical systems are described by the Von Neumann Hilbert space formulation. There exists a more general approach to quantum mechanics: the $C^*$-al\-ge\-bra\-ic formulation. It captures the Von Neumann formulation and in addition incorporates, amongst others, infinite-dimensional systems and systems with superselection rules~\footnote{As an example of an system with superselection rules, consider a quantum system consisting of fermions and bosons. A superselection rule forbids states which are superpositions of fermionic states and bosonic states.}. In the first instance the $C^*$-algebraic formulation was realized by Von Neumann, who wanted to generalize Pascual Jordan's work. Israel Gelfand, Mark Naimark and Irving Segal worked out the operator algebras of Von Neumann's and established the $C^*$-algebraic formulation of quantum mechanics.

\begin{postulate}[The $C^*$-al\-ge\-bra\-ic Postulate]
A physical system is described by a triplet $\{{\cal S}^*,{\cal A},<.;.>\}$, where: the observables are the Hermitian elements $\cal A$ of some unital $C^*$-algebra ${\cal B}$ called the algebra of observables; $S^*$ is the set of the possible states of the system, which is the collection of real-valued, positive linear functionals $\rho:{\cal A}\to \mathbb{C}$ satisfying $\rho(\mathbb{I})=1$; and $<.;.> : {\cal S}^*\times{\cal A}\mapsto \mathbb R$ is the prediction rule defined by $<\rho;A>\equiv\rho(A)$, corresponding to the expectation value of the observable $A\in{\cal A}$ when the system is in state $\rho\in {\cal S}^*$.
\end{postulate}

A $C^*$-algebra is defined formally as follows.

\begin{definition}
A $C^*$-algebra $\cal A$ is an involutive Banach algebra with the extra condition $||A^*A||=||A||^2$ for all $A\in \cal A$. 
\end{definition}

Involutive means that the algebra $\cal A$ is equipped with a $^*$-involution defined as a $\mathbb C$-antilinear map satisfying $(AB)^*=B^*A^*$ and $(A^*)^*=A$ with $A\in\cal A$. In the case that $\cal A=\B$, $^*$-involution is equal to normal Hermitian conjugation. A Banach algebra is an associative algebra over the complex or real numbers that is a Banach space as well, i.e. a complete, normed vector space satisfying $||AB||\leq ||A||\dot||B||$. This condition in particular implies that for elements $A$ of a $C^*$-algebra it holds that $||A||=||A^*||$.

By the GNS construction, standing for Gel'fand \& Naimark and Segal, every $C^*$-algebra is isomorphic to an algebra of bounded operators on some Hilbert space $\Hi$. This implies that the mathematical techniques of the Hilbert space formulation are still present in the $C^*$-algebraic language. So, in concreto, a $C^*$-algebra is a complex algebra of linear operators on a Hilbert space, closed in the norm topology of operators and closed under the involution (or conjugation) operator. 

Let us consider $\cal A=\B$. There exists for every physical state $\rho\in {\cal S}^*$, a corresponding density matrix $\hat{\rho}$ such that $\rho(A)=\mathrm{Tr}(\hat{\rho} A)$. It is clear that the $C^*$-al\-ge\-bra\-ic approach generalizes Von Neumann's approach, for it does not only capture $\Bsa$. For example, the observable algebra ${\cal A} = C(X)$, the algebra of all continuous functions on a metric space $X$, describes classical mechanics. 

In fact, a classical algebra is an Abelian or commutative algebra, all elements commute under the multiplication operation. A pure quantum algebra does not contain elements that commute with all other elements (except for the identity); the algebra is a factor, meaning that the intersection of the algebra $\cal A$ and its commutant ${\cal A}'$ (the elements commuting with $\cal A$) is
\begin{equation}
\cal A\cap \cal A' = \mathbb{C}\mathbb{I}_{\cal A}.
\end{equation}

\subsubsection{Quantum Mechanics As A Probability Theory}
An important aspect of quantum mechanics is its interpretation as a probability theory. In the $C^*$-algebraic approach this reveals itself evidently; states on commutative $C^*$-algebras induce probability measures via the Gel'fand transform and the Riesz representation theorem~\cite{Lan}. 

These theories prove that every state $\rho \in {\cal S}^*$ defines a regular positive measure $\mu_A$ on the Borel $\sigma$-algebra of the spectrum Spec($A$) of some $A \in \cal A$. This results in the definition of a functional $\mathbb E$ on functions on Spec($A$): $\mathbb{E}(f)\equiv\rho(f(A))=\int f(x)\mu_A(dx)$. The suggestive notation of this functional $\mathbb E$, leads one to suspect that it is interpreted as the expectation value of $f(A)$.

\subsection{Quantum Mechanics In This Thesis}
For the sake of generality, I will work within the $C^*$-algebraic formulation. However, because I only consider finite-dimensional systems, I can make use of Hilbert space techniques. Let $d$ be the dimension of a finite-dimensional system. The Hilbert space that describes this system is denoted by the complex vector space $\Hi=\C$. The observables $\Bsa$ are the Hermitian elements of the complex $d\times d$ matrices. The states are described by positive Hermitian $d\times d$ matrices $\hat{\rho}$ normalized with $\mathrm{Tr}(\hat{\rho})=1$. More on density matrices is found in section~\ref{sec:dens}.

\newpage
\section{Quantum Operations}
\label{sec:qo}
Quantum systems interact with their environment. Interaction can be seen as the processing of information from one system to another and can be both of quantum and of classical nature. Maps that describe the interaction are called quantum operations.

In defining quantum operations, it is important to stress the difference between the Schr\"odinger and Heisenberg \emph{pictures}. 

In the Schr\"odinger picture an operation $T^*$ is a map taking states on a system with observable algebra $\cal A$ to states on a system with an algebra of observables $\cal B$. Since the set of states $S^*$ is a subset of the dual of the algebra of observables, $S^*\subseteq A^*$, an operation $T^*$ on states maps the \emph{dual} $\cal A^*$ of $\cal A$ to the dual $\cal B^*$ of $\cal B$:
\begin{equation}
T^*: {\cal A}^* \to {\cal B}^*.
\end{equation}

In the Heisenberg picture, the action of an operation $T$ is characterized by the way it influences measurement of observables~\cite{Wer}. Measurement of an observable $B$ is obtained by application of an operation that takes a system with algebra of observables $\cal A$ to a system with algebra of observables $\cal B$. First apply the channel, then measure the observable $B$. This is effectively measurement on the system with algebra of observables $\cal A$ and is denoted by $T(B)$:

\begin{equation}
T: {\cal B} \to {\cal A}.
\end{equation}

The operations $T:\cal B\to \cal A$ and $T^*:\cal A^* \to \cal B^*$ are related by
\begin{equation}
(T^*(\rho))(B)=\rho(T(B))
\end{equation}
in which $\rho\in A^*$ is a state on the system with observable algebra $\cal A$. The action of an operation $T$ on a density operator $\hat\rho$ is written as $T^*(\hat\rho)$. Notice that $\mathrm{Tr}(T^*(\hat\rho)a)=\mathrm{Tr}(\hat\rho T(a))$.

A definition of quantum operations is attained by contemplation on the conditions laid down by quantum mechanics. First of all, a quantum operation has to be linear in its arguments, for it has to cover action on mixtures of states. Then, for the fact that is maps states to states, it has to be positive and unit-preserving. At last, the action of $\mathrm{id}_n\otimes T$ on just part of a composite system ${\cal M}_n \otimes {\cal B}$, has to be positive for all n-dimensional systems ${\cal M}_n$ as well. This non-trivial requirement is called complete positivity.

\begin{definition}[Quantum Operation]
A quantum operation converting a system with observable algebra $\cal A$ to a system with observable algebra $\cal B$ is a completely positive (CP), unit-preserving, linear map $T : {\cal B}\to \cal A$.
\end{definition}

\subsection{Quantum Dynamics And Quantum Operations}
A symmetry of a quantum system is by definition a bijection onto itself, or consequently an automorphism. In the case $\cal A=\B$, the symmetries of quantum systems are unitary implemented maps $T(A)=UAU^*$. As noted for the dynamics in Von Neumann's Hilbert space formulation (see section~\ref{sec:qm}), Wigner's theorem not only justifies the unitary implemented maps, it also states that symmetries can be of the form $T(A)=WAW^*$ with $W$ an anti-unitary operator. However, because operations implemented by anti-unitary operators are not completely positive, they can only act on global systems and make no sense on subsystems. Thereby, time-reversal or spin flipping operations, which are implemented by anti-unitary operators, are in general not possible.

\subsection{Heisenberg Principle For Quantum Operations}
The Heisenberg principle applies not only to measurement, it is significant for quantum operations in general. From this point of view, the Heisenberg principle states that is impossible to transfer quantum information from one system to another without distortion. 

\begin{theorem}[Heisenberg Principle]
Let $T:\cal A\otimes \cal B \to\cal A$ be a quantum operation satisfying
\begin{equation}
T(a\otimes \mathbb{I})=a
\end{equation}
for all $a\in\cal A$. Then
\begin{equation}
T(\mathbb{I}\otimes b)\in \cal A\cap \cal A'.
\end{equation}
\end{theorem}
This implies that if $\cal A$ describes a pure quantum system and thus its centre is $\cal A\cap \cal A' = \mathbb{C}\mathbb{I}_{\cal A}$, then 
\begin{equation}
b\mapsto T(\mathbb{I}\otimes b)=z(b)\mathbb{I}_{\cal A},\qquad b\in\cal B
\end{equation}
with $z(b)\in \mathbb{C}$; if the system is totally quantum, then this non-distorting operation $T$ has not transferred information at all. 

\subsection{Stinespring Dilation Theorem}
The following theorem is known as the Stinespring dilation theorem. It connects unitary-implemented maps known from standard quantum mechanics with quantum operations as defined above.
\begin{theorem}[Stinespring]
\label{the:sti}
Let $\cal A$ be a unital $C^*$-algebra and let $T:{\cal A}\to {\cal B} \subset {\cal B(K)}$ be a CP map. Then there exist a Hilbert space $\cal H'$, a bounded operator $V:{\cal K}\to {\cal H'}$, and a $^*$-homomorphism $\pi : {\cal A}\to {\cal B(H')}$ such that for all $a\in {\cal A}$:
\begin{equation}
T(a)=V^* \pi (a) V.
\end{equation}
\begin{displaymath}
\xymatrix{
{\cal A} \ar[dr]_\pi \ar[r]^T & {\cal B} \subset {\cal B(K)}\\
& {\cal B(H')} \ar[u]_{V^*\cdot V}}
\end{displaymath}
Up to unitary transformations there is only one choice of $({\cal H'},V,\pi)$ (called the Stinespring dilation) such that the vectors $\pi(a)V\phi$ generate ${\cal H'}$. If $T(\mathbb{I})=\mathbb{I}$ (T is a quantum operation), then $V$ is an isometry, i.e. $V^*V=\mathbb{I}$.
\end{theorem}

As a special case of the Stinespring dilation theorem, consider a CP map $T :{\cal B(H)}\to{\cal B(K)}$. In this case ${\cal H'}$ is given by ${\cal H}'={\cal H}\otimes{\cal E}$ and $V : {\cal K} \to {\cal H}\otimes{\cal E}$ is such that
\begin{equation}
T(a)=V^* a \otimes \mathbb{I}_{{\cal E}}V\qquad \forall a\in {\cal B(H)}.
\end{equation}
This is due to the fact that a normal $^*$-representation of the $C^*$-algebra $\B$ is unitarily equivalent to the amplification map $a \mapsto a \otimes \mathbb{I}_{{\cal E}}$. See~\cite{Rag}. 

\begin{figure}[!htb]
\begin{center}
\includegraphics[width=8cm]{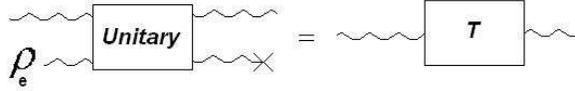}
\caption{The Stinespring dilation.}
\label{fig:stine}
\end{center}
\end{figure}

For physicists a particular form of the Stinespring dilation is known as the operator-sum representation. In the Schr\"odinger picture, this representation is constructed as illustrated in fig.~(\ref{fig:stine}). An initial quantum system is first coupled to an ancillary system, i.e. its environment, and followed by unitary evolution of the composite system. At the end, the environment is disregarded. So the Stinespring dilation theorem in this form states that every quantum operation is given by
\begin{equation}
T^*(\hat\rho)=\mathrm{Tr}_\mathrm{env}(U\hat\rho\otimes\hat\rho_\mathrm{env}U^*)
\end{equation}
in which the subscript env denotes the environment and $U$ is a unitary operator. In the Heisenberg picture this is equivalent to
\begin{equation}
T(a)=U^*a\otimes\mathbb{I}_\mathrm{env}U.
\end{equation}

In this form, the Stinespring dilation theorem is the mathematical foundation of the idea that any evolution of quantum systems is implemented by unitaries, as assumed in the Copenhagen interpretation of quantum mechanics. Realize that the motion of the quantum system if seen uncoupled to the environment, may not be unitary. 

Stinespring's theorem connects a CP map with its Kraus representation
\begin{equation}
T(a)=\sum_i K_i^*aK_i,
\end{equation}
in which $K_i$ are bounded operators, called Kraus operators. Kraus operators satisfy
\begin{equation}
\sum_i K_i^*K_i =\mathbb{I}
\end{equation}
for quantum operations (trace-preserving CP maps). The Kraus representation of a CP map from its Stinespring dilation is obtained in the following way~\footnote{In fact, all Kraus representations are constructed like this. This is a consequence of a Radon-Nikodym-like theorem. See Ref~\cite{Rag}.}. Let $\sum_i\ket{\psi_i}\bra{\psi_i}=\mathbb{I}$. Then
\begin{align}
T(a)&=V^*a\otimes \mathbb{I}V\nonumber\\
&=\sum_i V^*a\otimes\ket{\psi_i}\bra{\psi_i}V\nonumber\\
&\equiv\sum_i\tilde V_i^*a\tilde V_i,
\end{align}
where $\tilde V_i$ are bounded operators, satisfying $\sum_i \tilde V_i^*\tilde V_i =\mathbb{I}$. If the dimension of the system is $d$, the minimal dilation consists of a maximal number of $d^2$ of Kraus operators (or Stinespring operators).

\newpage
\section{Impossible Operations: Quantum Cloning}
\label{sec:noclone}
An important theorem in quantum theory is the no-cloning theorem. It states that perfect cloning of a quantum system is impossible. It is a fundamental theorem with deep impact in quantum information theory. First, I will give a formulation of the no-cloning theorem in terms of quantum information and CP maps. Then I will discuss the theorem in a more physical setting; the setting in which it was first discovered. 

\subsection{No-Cloning Theorem}
\begin{figure}[!htb]
\begin{center}
\includegraphics[width=8cm]{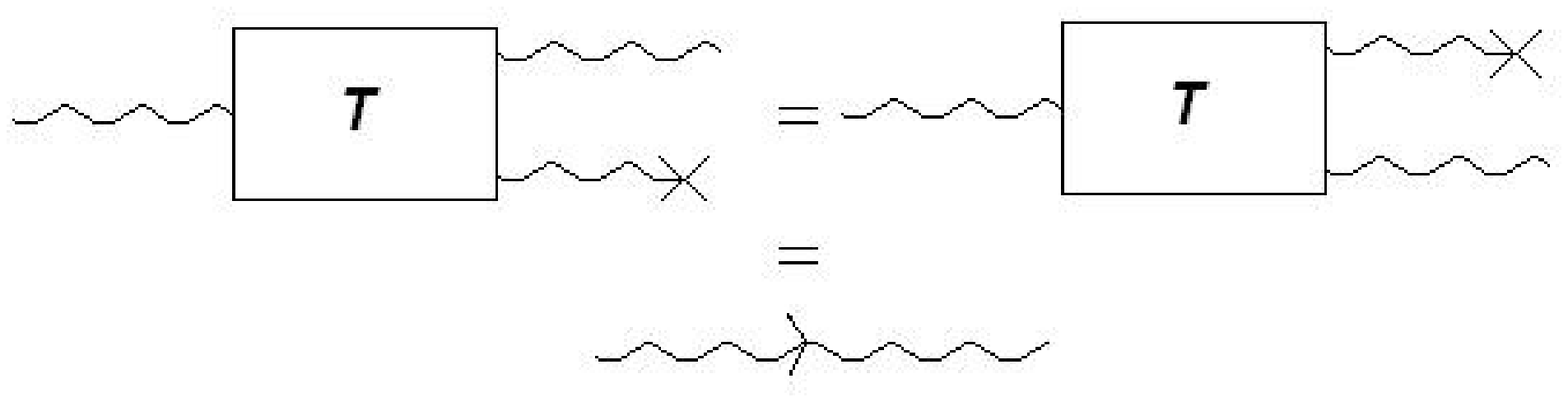}
\caption{A perfect quantum cloner.}
\label{fig:clone}
\end{center}
\end{figure}

A symmetric cloning machine $T$ is a machine that makes a perfect copy of some arbitrary unknown quantum state. If we would throw one of the copies away, we would have a state that is identical to the input state. Fig.~(\ref{fig:clone}) is an illustration of such a device.

Let $\cal A$ be the observable algebra of the system to be cloned. The cloning operation is expressed in the Heisenberg picture by
\begin{equation}
T(a\otimes\mathbb{I})=T(\mathbb{I}\otimes a)=a
\end{equation}
with $a\in\cal A$. 

The no-cloning theorem forbids such machines in the case that $\cal A$ is non-Abelian, for instance a pure quantum algebra.

\begin{theorem}[No-Cloning Theorem]
Let $T:\cal A \to \cal A$ be a quantum operation. If
\[
T(a\otimes\mathbb{I})=T(\mathbb{I}\otimes a)=a,
\]
then $\cal A$ is Abelian.
\end{theorem}

Note that as a corollary, only classical (central) information can be extracted from a quantum system without distortion.
 
In the following section, I will show how this impossibility of quantum cloning was found in a setting of quantum optics. This section might be considered as standing on its own in this thesis, and in fact it is. Just think of it as nice example of the interplay between physics and mathematics.
 
\subsection{Wootters' and Zurek's No-Cloning Theorem}
\emph{A single quantum cannot be cloned} is the title of an important Letter to Nature by Wootters and Zurek~\cite{Woot} in 1982. Their notion of the impossibility of quantum cloning is now considered as the no-cloning theorem. Although Wootters and Zurek originally stressed this impossibility in the framework of quantum optics, the no-cloning theorem is a fundamental theorem that forbids \emph{perfect} copying of an arbitrary state of a quantum system. In fact, it is a direct consequence of quantum mechanics and one of its manifestations is the prohibition of superluminal communication. 

Consider a single photon, that can be polarized horizontally $\ket{\rightarrow}$ or vertically $\ket{\uparrow}$. The operation of perfect quantum cloning should have the following effect on the states $\ket{\uparrow}$ and $\ket{\rightarrow}$:
\begin{equation}
\ket{A_0}\ket{\uparrow}\qquad\longrightarrow\qquad\ket{A_\mathrm{hor}}\ket{\upuparrows}
\end{equation}
and
\begin{equation}
\ket{A_0}\ket{\rightarrow}\qquad\longrightarrow\qquad\ket{A_\mathrm{vert}}\ket{\rightrightarrows}.
\end{equation}

In these equations, $\ket{A_0}$, $\ket{A_\mathrm{hor}}$ and $\ket{A_\mathrm{vert}}$ refer to the the states of the cloning machine, before cloning, after cloning of a horizontally polarized photon and after cloning of a vertically polarized photon respectively. The symbols $\ket{\upuparrows}$ and $\ket{\rightrightarrows}$ represent the states of the radiation field in which there are two photons, that are both polarized horizontally or both polarized vertically. 

Operations on quantum mechanical systems are by assumption implemented by linear and in fact, unitary operators. In addition, states are allowed that are superpositions of eigenstates of some observable, in this case superpositions of $\ket{\rightarrow}$ and $\ket{\uparrow}$. It follows that by linearity a perfect cloning machine should affect the superposition state $\alpha\ket{\uparrow}+\beta\ket{\rightarrow}$ as
\begin{equation}
\label{eq:lincm}
\ket{A_0}(\alpha\ket{\uparrow}+\beta\ket{\rightarrow})	\qquad\longrightarrow\qquad
\alpha\ket{A_\mathrm{vert}}\ket{\upuparrows} + \beta\ket{A_\mathrm{hor}}\ket{\rightrightarrows}.
\end{equation}
If the machine is \emph{universal}, i.e. the states $\ket{A_\mathrm{hor}}$ and $\ket{A_\mathrm{vert}}$ are the same, the photons are in the pure state
\begin{equation}
\alpha\ket{\upuparrows} + \beta\ket{\rightrightarrows}.
\end{equation}
In the non-universal case, the photons are in a mixed state. However in both cases, these states are not the same as state in which both photons are in the superposition state $\alpha\ket{\uparrow}+\beta\ket{\rightarrow}$. Let $\ket{0}$ be the vacuum state and let $a^\dagger_\mathrm{vert}$ and $a^\dagger_\mathrm{hor}$ be raising operators. The state in which both photons are in the superposition state $\alpha\ket{\uparrow}+\beta\ket{\rightarrow}$ is given by
\begin{equation}
2^{-1/2}(\alpha a^\dagger_\mathrm{vert} + \beta a^\dagger_\mathrm{hor})^{\otimes 2}\ket{0}=\alpha^2\ket{\upuparrows} + 2^{1/2}\alpha\beta\ket{\uparrow\rightarrow}+\beta^2\ket{\rightrightarrows}
\end{equation}
which is not the same as the state in eq.~(\ref{eq:lincm}), neither in the universal nor the non-universal case. This proves the no-cloning theorem. As said above, this theorem does not prohibit the perfect cloning of some states, for example the cloning of $\ket{\rightarrow}$ and $\ket{\uparrow}$, it states that it is impossible to clone an arbitrary state of a quantum system. Naturally, the validity of the no-cloning theorem is not restricted to cloning of polarization states. The same argument used here can be extended to any quantum system of arbitrary dimension. 

If perfect quantum cloning were possible, it would mean the offending of Einstein's special relativity. Consider a pair of entangled spin-$1/2$ particles (or Einstein-Podolsky-Rosen pairs of photons), such that measurement on one of the members of the pair fixes the state of the other one, that may be far away. If, before measurement of the first particle, the owner of the second particle could have made infinitely many perfect clones of his particle and thus would have known exactly the state of this particle by statistical estimation, he could say with infinite accuracy what measurement was made on the first member of the original pair. If this is done within the time that light needs to travel from the first to the second observer, the not-faster-than-light axiom is violated and superluminal communication becomes available.

\subsubsection*{Optimal cloning}
Although perfect cloning is not possible, the search for the optimal, i.e. as good as possible, cloning machines is interesting. Research in this area has produced many explicit boundaries for several quantum cloning schemes, such as universal pure state cloning~\cite{Wer2} and phase covariant pure state cloning~\cite{Dar}. 

Most important is an article by Werner and Keyl~\cite{Wer2} on optimal cloning of pure states. Consider a universal (i.e. no discrimination between input states) pure state quantum cloning machine $T$, that copies $N$ identically prepared input states to $M$ optimal copies, which are of course not perfect copies. Werner found a bound on the accuracy of this cloning device in terms of the fidelity $F$ (see for details section~\ref{sec:fid}). This fidelity of a quantum cloner is defined by
\begin{equation}
F=\braket{\psi}{\rho_\mathrm{out}|\psi}
\end{equation}
in which $\rho_\mathrm{out}\equiv\mathrm{Tr}_{M-1}(T(\ket{\psi}\bra{\psi}^{\otimes N}))$ is the reduced density matrix of one of the clones ($\mathrm{Tr}_{M-1}$ is a partial trace over $M-1$ clones). The fidelity is thus the probability overlap between one of the $N$ unknown input states and one of the imperfect copies $M$~\footnote{Only one of the clones is compared to an input clone, because correlations between clones may increase the value of our figure of merit misleadingly. It would give us a false idea about the quality of the cloner, since a cloner has to copy uncorrelated clones (by definition). However Ref~\cite{Wer2} shows that this judging of single clones yields the same fidelity as the fidelity between all $M$ imperfect copies and $M$ (hypothetical) perfect copies. This implies that the optimal cloner, produces uncorrelated clones. See for details~\cite{Wer1}.}. The upper bound is given by~\cite{Wer2}
\begin{equation}
F_\mathrm{opt}=\frac{N}{M} + \frac{M-N}{M}\frac{N+1}{d+N}
\end{equation}
in which $d$ is the dimension of the system in consideration. While the polarization of a photon is two-dimensional and when the setting is restricted to one input particle, this equation reduces to
\begin{equation}
F_\mathrm{opt}=\frac{2}{3} + \frac{1}{3M}.
\end{equation}
Note that if $M\to\infty$, then $F_\mathrm{opt}=\frac{2}{3}$ which is the maximal fidelity obtained by optimal measurement of a single quantum system (qubit). To see this, notice that measuring a qubit (in a pure state) and preparing infinitely many clones according to the outcome of the measurement is equivalent to cloning of infinitely many clones~\cite{Bru}. This fidelity of course can never be $F^\mathrm{meas}=1$, for this would imply that one measurement would give an outcome that is fully accurate. Equivalently, the state of infinitely many clones can be estimated precisely (statistically), and if the fidelity of the clones were $1$, the state of the original qubit would be known. This cannot be true either. 

However, the limit formula $F^\mathrm{meas}_\mathrm{opt}=\lim_{M\to\infty} F_\mathrm{opt}(1,M)$ has not yet been proved to hold in all case. It is not trivial, since correlations and entanglement between clones cannot be neglected. Nevertheless, in the case of phase-covariant cloners\cite{Bru2} and of pure state cloners~\cite{Wer2}, the formula is true. 

\subsubsection*{Quantum Cloning and Stimulated Emission}
In quantum cloning of polarization states, it is important to note that perfect cloning in a framework of stimulated emission is not possible due to perturbation by spontaneous emission~\cite{Man,Mil}. Consider a quantum cloner based on stimulated emission, i.e. an amplifier. Let $a$ and $b$ be two resonant planes of an excited $3$-level atom with orthogonal transition dipole moments $\vec \mu_a=|\mu|\vec\epsilon_a$ and $\vec \mu_b=|\mu|\vec \epsilon_b$ in which $\vec \epsilon_{a,b}$ are two orthogonal unit polarization vectors. The input state of the composite system is given by $\ket{1_{\epsilon_1},0_{\epsilon_2}}\ket{+_a,+_b}$ with $\ket{1_{\epsilon_1},0_{\epsilon_2}}$ the initial state of the field with one photon polarized in direction $\vec \epsilon_1$ and $\ket{+_a,+_b}$ the state of the two excited atoms. The state of the system after interaction with the photon is given by
\begin{equation}
\ket{\psi_\mathrm{f}}=\mathrm{exp}(-i\hat{H}_I\Delta t/\hbar)\ket{1_{\epsilon_1},0_{\epsilon_2}}\ket{+_a,+_b}
\end{equation}
with the electric dipole interaction Hamiltonian
\begin{equation}
\hat{H}_I=g\sum_{s=1}^2(\hat\sigma_a^{(-)}\vec \mu_a + \hat\sigma_b^{(-)}\vec \mu_b)\cdot\vec \epsilon_s^* \hat a_s^\dagger + h.c.
\end{equation}
where $g$ is a coupling constant, the dot stands for the normal complex inner product on a two dimensional Hilbert space and $\hat \sigma$ and $\hat a$ denote the atomic and field lowering and raising operators for the different modes. For short times $\Delta t$, a Taylor expansion of the time evolution operator can be made. The zeroth-order term corresponds to no interaction, i.e. $\ket{\psi_f}=\ket{\psi_i}$. So, sometimes this operation is not a cloning operation at all. The first-order term leads to the unnormalized state
\begin{align}
\ket{\psi_\mathrm{f}}&=\ket{-_a,+_b}\{\sqrt{2}\vec \mu_a\cdot\vec \epsilon_1^*\ket{2_{\epsilon_1},0_{\epsilon_2}}+ \vec \mu_a\cdot\vec \epsilon_2^*\ket{1_{\epsilon_1},1_{\epsilon_2}}\}\nonumber\\
&+\ket{+_a,-_b}\{\sqrt{2}\vec \mu_b\cdot\vec \epsilon_1^*\ket{2_{\epsilon_1},0_{\epsilon_2}}+ \vec \mu_b\cdot\vec \epsilon_2^*\ket{1_{\epsilon_1},1_{\epsilon_2}}\}.
\end{align}
Tracing over the atomic variables yields the normalized density operator
\begin{equation}
\hat \rho = \frac{2}{3}\ket{2_{\epsilon_1},0_{\epsilon_2}}\bra{2_{\epsilon_1},0_{\epsilon_2}}+\frac{1}{3}\ket{1_{\epsilon_1},1_{\epsilon_2}}\bra{1_{\epsilon_1},1_{\epsilon_2}}
\end{equation}
which is a mixed two-photon state. The first term in this expression corresponds to stimulated emission and thus to the production of two clones (so an extra photon besides the original one, since both photons are clones). The second term is attributable to spontaneous emission, since the polarization of spontaneous emission is arbitrary. Note that the probability that the input state is cloned is twice the probability that an anti-clone, i.e. orthogonal to the initial state, is produced. Since the fidelity, i.e. the probability overlap between input state and output state, is the relative frequency of photons of the right polarization in the final state, it is clear that in this case the fidelity is given by 
\begin{equation}
F=\frac{2}{3}\times 1 + \frac{1}{3}\times \frac{1}{2}=\frac{5}{6}.
\end{equation}
Namely with a probability of $\frac{2}{3}$ both clones are equal to the initial state and with a probability of $\frac{1}{3}$ only one of the clones is equal to the initial one, such that in this case there is a chance of $\frac{1}{2}$ to pick a right clone. As said above, a fidelity of $\frac{5}{6}$ was found to be optimal.
\newpage
\section{Mathematical Tools}
\label{sec:mt}

\subsection{Density Operators}
\label{sec:dens}
In the last chapter of this thesis, I will need to do some explicit calculation on density operators. Therefore a closer look on the set of density operators is necessary.

Let $\ket{e_x} \in \Hi$ be an orthonormal basis of a finite-dimensional Hilbert space $\Hi$. In this basis the matrices $e_{xy}=\ket{e_x}\bra{e_y}$ form a basis of ${\cal B(H)}$. Now the density matrix $\hat\rho$ is defined by $\hat\rho=\sum_{xy}\rho_{xy}e_{xy}$ in which the expansion coefficients $\rho_{xy}$ are given by $\rho_{xy}=\mathrm{Tr}(\rho e_{xy})=\braket{e_y}{\rho e_x}$. 

The expectation value of an observable $A$ is given by 
\begin{equation}
\rho(A)=\mathrm{Tr}(\hat\rho A)
\end{equation}
where $\rho_ii$ are non-negative and normalized (summed to one) and thus interpreted as probabilities.  

As an example, consider a state on a two-dimensional system, $d=2$. It has to be a positive, Hermitian $2\times 2$ matrix with $\mathrm{Tr}(\hat\rho)=1$, i.e. it is written as
\begin{align}
\hat\rho &= \frac{1}{2}\left(\begin{array}{c c}
1 + r_3& r_1-ir_2 \\
r_1-ir_2 & 1-r_3\end{array}\right)=\frac{1}{2}(\mathbb{I}+\vec{\sigma}\cdot\vec{r})\\
r_i&=\mathrm{Tr}(\hat\rho \sigma_i)\nonumber
\end{align}
in which $\vec{\sigma}=(\sigma_x,\sigma_y,\sigma_z)$ are the Pauli matrices. The vector $\vec{r}$ is called the Bloch vector and because of positivity of $\rho$ satisfies $|\vec{r}|^2\leq 1$. Pure states satisfy the extra condition $\hat\rho^2=\hat\rho$, such that $|\vec{r}|^2 = 1$. Mixed states have $|\vec{r}|^2< 1$. The sphere of pure states is called the Bloch sphere. The rotation invariant Haar measure on the space of pure states for $d=2$ is thus given by the Haar measure on the unit-sphere ${\cal S}^2$ in three-dimensional Euclidean space.

\subsection{Projective Hilbert Space}
The space of all pure states of $\mathbb{C}^2$ or in other words all one-dimensional projections, is called projective Hilbert space and is denoted by $\PO$. As said above for $d=2$ the set of pure states corresponds to $\PO\cong {\cal S}^2$. Although there exists a generalized Bloch sphere representation for $d$-dimensional systems,
\begin{equation}
\frac{1}{d}\mathbb{I}_d + \frac{1}{2}\sum_{i = 1}^{d^2 -1}r_i\lambda_i,
\end{equation}
with $r_i \in \mathbb{R}$ and $\lambda_i$ the infinitesimal generators of the group $SU(d)$, which together with $\mathbb{I}_d$ span the Hermitian $d\times d$ matrices, it is unfortunately not true that the Bloch vectors $\vec{\lambda}=(\lambda_1,\dots,\lambda_{d^2-1})$ form a $d^2-1$ unit-ball. See for details~\cite{Kim}. Therefore, the invariant Haar measure on $\PH$ has to be calculated explicitly~\footnote{The following section is a partly review of Ref~\cite{Cav}.}. 

\subsubsection{The Fubini-Study Metric}
\begin{theorem}[Haar Measure On Projective Hilbert Space]
Let $\C$ be the Hil\-bert space describing a system of dimension $d$. Then the rotation invariant Haar measure $\D p$ on the projective space, $\PH$, is given by
\begin{equation}
\D p=\sin^{2d-3} \theta\cos \theta \D\theta \D{\cal S}_{2d-3},\qquad 0\leq \theta \leq \pi / 2,
\end{equation}
in which $\D {\cal S}_{2d-3}$ denotes the Haar measure on a Euclidean $2d-3$-dimensional unit-ball.
\end{theorem}
\emph{Proof:} Consider two normalized (pure) states $\ket{\psi}$ and $\ket{\psi}+\ket{\D\psi}$ on the Hilbert space $\Hi$ that are close to each other. The infinitesimal angle $\D s$ between these two states satisfies
\begin{align}
\cos^2(\D s)=1-\D s^2&=|\bra{\psi}(\ket{\psi}+\ket{\D\psi})|^2\nonumber\\
&=1+2\Re(\braket{\psi}{\D\psi})+|\braket{\D\psi}{\D\psi}|^2.
\end{align}
Because of normalization,
\begin{eqnarray}
0=\D(\braket{\psi}{\psi})=\braket{\D\psi}{\psi}+\braket{\psi}{\D\psi}+\braket{\D\psi}{\D\psi}=2\Re(\braket{\psi}{\D\psi})+\braket{\D\psi}{\D\psi}\nonumber\\
\Rightarrow 2\Re(\braket{\psi}{\D\psi})=-\braket{\D\psi}{\D\psi}
\end{eqnarray}
which is valid up for first order in small displacements, the angle $\D s$ is thus given by
\begin{equation}
\D s^2=\braket{\D\psi}{\D\psi}-|\braket{\psi}{\D\psi}|^2
\end{equation}

This metric is called the Fubini-Study metric. It is invariant under phase changes of $\ket{\psi}$ and $\ket{\psi}+\ket{\D\psi}$. Now, let two orthogonal vectors $\ket{\psi_0}$ and $\ket{\eta}$ decompose the state $\ket{\psi}$ as
\begin{equation}
\ket{\psi}=e^{i\phi}\cos\theta\ket{\psi_0}+\sin\theta\ket{\eta}
\end{equation}
with $0\leq\theta\leq \pi/2$. Because $\D s$ is invariant under phase changes, the phase factor $\phi$ is put equal to $\phi=0$. The vector $\ket\eta$ is a vector in the plane orthogonal to $\ket{ \psi_0}$. Clearly, 
\begin{equation}
\ket{\D\psi}=-\sin\theta \D\theta\ket{\psi_0}+\cos\theta \D\theta\ket{\eta}+\sin\theta\ket{\D\eta},
\end{equation}
such that
\begin{align}
\D s^2&=\braket{\D\psi}{\D\psi}-|\braket{\psi}{\D\psi}|^2\nonumber\\
&=\D\theta^2+\sin^2\theta\braket{\D\eta}{\eta}-\sin^4\theta|\braket{\eta}{\D\eta}|^2\nonumber\\
&=\D\theta^2+\sin^2\theta \D\gamma^2
\end{align}
with $\D\gamma^2=\braket{\D\eta}{\eta}-\sin^2\theta|\braket{\eta}{\D\eta}|^2$. In this equation, I used $\braket{\psi_0}{\D\eta}=0$ and again because of normalization,
\begin{equation}
\Rightarrow 2\Re(\braket{\eta}{\D\eta})=-\braket{\D\eta}{\D\eta}.
\end{equation}
The metric $\D\gamma^2$ defines a Riemannian metric on the space of normalized vectors orthogonal to $\ket{\psi_0}$. This is a subspace of the space orthogonal to $\psi_0$, which itself is $d-1$-dimensional. The normalized vectors form a Euclidean unit-ball of dimension $2(d-1)-1=2d-3$ denoted by ${\cal S}_{2d-3}$. The line element $\D\gamma^2$ differs from the normal geometry on a $(2d-3)$-dimensional unit-sphere. 

In order to see this, fill $\ket\eta$ up with $d-2$ vectors $\{\ket{\eta_{j=2,\ldots, d - 1}}\}$ such that the set $\{\ket\eta,\ket{\eta_{j=2,\ldots, d - 1}}\}$ forms an orthogonal basis on the complex $d-1$-dimensional subspace, orthogonal to $\ket{\psi_0}$. In this basis an arbitrary vector $\ket{\eta'}$ is decomposed as
\begin{equation}
\ket{\eta'}=(x_1+iy_1)\ket{\eta}+\sum_{j=2}^{d-1}(x_j+iy_j)\ket{\eta_j}
\end{equation}
and normalization implying
\begin{equation}
\sum_{j=1}^{d-1}x_j^2+y_j^2=1.
\end{equation}

The first term of the line element $\D\gamma^2$ is now given by
\begin{equation}
\braket{\D\eta}{\D\eta}=\sum_{j=1}^{d-1}(\D x_j^2+\D y_j^2).
\end{equation}
and the second term by
\begin{equation}
-\sin^2\theta|\braket{\eta}{\D\eta}|^2=-\sin^2\theta(\D x_1^2+\D y_1^2).
\end{equation}
Since 
\begin{equation}
\D\sum_{j=1}^{d-1}x_j^2+y_j^2=\sum_{j=1}^{d-1}x_j\D x_j+y_j\D y_j=0
\end{equation}
and
\begin{equation}
\D x_1=\frac{1}{x_1}\left(-y_1\D y_1-\sum_{j=2}^{d-1}x_j^2+y_j^2\right)=0,
\end{equation}
this implies that the line element $\D\gamma^2$ is given by
\begin{align}
\D\gamma^2&=\braket{\D\eta}{\eta}-\sin^2\theta|\braket{\eta}{\D\eta}|^2\nonumber\\
&=\sum_{j=2}^{d-1}(\D x_j^2+\D y_j^2)+\D y_1^2-\sin^2\theta \D y_1^2\nonumber\\
&=\sum_{j=2}^{d-1}(\D x_j^2+\D y_j^2)+\cos^2\theta \D y_1^2
\end{align}
It turns out that the Haar measure on ${\cal S}_{2d-3}$ at $\ket{\eta}$, defined by the metric $\D\gamma^2$, is
\begin{equation}
\cos\theta \D y_1\D x_2\ldots \D x_{d-1}\D y_2\ldots \D y_{d-1}=\cos\theta \D {\cal S}_{2d-3}
\end{equation}
in which $\D {\cal S}_{2d-3}$ denotes the normal Haar measure on a Euclidean $(2d-3)$-dimensional unit-sphere. As expressed by the Fubini-Study metric $\D s^2$, all lengths on ${\cal S}_{2d-3}$ are scaled with a factor $\sin \theta$, and it follows that the Haar measure on the projective Hilbert space is expressed by
\begin{equation}
\D p=\sin^{2d-3} \theta\cos \theta \D\theta \D {\cal S}_{2d-3}
\end{equation}
with $0\leq \theta \leq \pi / 2$.\begin{flushright}$\Box$\end{flushright}

\subsection{Fidelity}
\label{sec:fid}
As mentioned in the introduction, this thesis is about the Heisenberg principle. Therefore, it may not be a suprise that a notion of quality of quantum operations is needed. After all, the amount of information extraction or distortion is to be captured and compared. An appropriate figure of merit is the \emph{fidelity}. I already used the fidelity in section~\ref{sec:noclone} and defined it loosely as the overlap probability between two pure states. Here, I will give a more formal defintion and prove some properties of the fidelity.

\begin{definition}[Fidelity]
Let $\hat \rho$ and $\hat \sigma$ be two density operators (pure or mixed). The \emph{fidelity} $F$ of $\hat \rho$ and $\hat \sigma$ is 
\begin{equation}
\label{eq:fid}
F(\hat \rho, \hat \sigma)=\left(\mathrm{Tr}(\sqrt{\hat \rho^{1/2} \hat \sigma\hat \rho^{1/2}})\right)^2.
\end{equation}
\end{definition}

If $\hat \rho$ and $\hat \sigma$ are pure states, i.e. $\hat \rho=\ket\psi\bra\psi$ and $\hat\sigma=\ket\phi\bra\phi$, then the fidelity $F$ reduces to the pure state fidelity.

\begin{definition}[Pure State Fidelity]
Let $\hat \rho=\ket\psi\bra\psi$ and $\hat\sigma=\ket\phi\bra\phi$ be two pure density operators. The \emph{pure state fidelity} $F$ of $\ket\psi$ and $\ket\phi$ is 
\begin{equation}
F(\ket\psi,\ket\phi)=|\braket{\psi}{\phi}|^2.
\end{equation}
\end{definition}

The pure state fidelity is thus the overlap probability between two pure quantum states. 

The following three theorems characterize fidelity. See Ref~\cite{Niels} for more on fidelity.
\begin{theorem}[Uhlmann's Theorem] 
\label{the:uhl}
Let $\hat \rho$ and $\hat \sigma$ be two density operators. Then the fidelity (eq.~(\ref{eq:fid})) is equal to
\begin{equation}
F(\hat \rho, \hat \sigma)=\mathrm{max}_{\ket\psi,\ket\phi}|\braket{\psi}{\phi}|^2
\end{equation}
in which the maximization is over all \emph{purifications} $\ket{\psi}$ of $\hat\rho$ and $\ket\phi$ of $\hat\sigma$.
\end{theorem}
In this definition the concept of purification is used. Suppose that $\hat\rho=\sum_i p_i\ket{i}\bra{i}$ is the orthonormal decomposition of $\hat\rho$. Define the pure state $\ket{\psi}$ by
\begin{equation}
\ket\psi=\sum_i\sqrt{p_i}\ket{i}\ket{i_a}
\end{equation}
where $\ket{i_a}$ is a vector state on a ancilla quantum system $a$ described by a state space identical to the initial system. It is easy to see that restriction of the pure state $\ket\psi$ to the initial system is exactly our initial (mixed) quantum state $\hat\rho$:
\begin{align}
\mathrm{Tr}_a(\ket\psi\bra\psi)&=\sum_{ij}\sqrt{p_ip_j}\ket{i}\bra{j}\mathrm{Tr}(\ket{i_a}\bra{j_a})\nonumber\\
&=\sum_i p_i\ket{i}\bra{i}\nonumber\\
&=\hat\rho.
\end{align}
The pure state $\ket\psi$ is called the purification of $\hat\rho$. The proof of theorem~\ref{the:uhl} can be found in Ref~\cite{Niels}.

The following theorem states that a quantum operation cannot improve the distinguishability of two quantum states.

\begin{theorem}[Monotonicity Of The Fidelity]
Let $T^*$ be a quantum operation. Then 
\begin{equation}
F(T^*(\hat\rho),T^*(\hat\sigma))\geq F(\hat\rho,\hat\sigma).
\end{equation}
\end{theorem}
\emph{Proof:} Let $\ket\psi$ be the purification of $\hat\rho$ and $\ket\phi$ of $\hat\sigma$ and let $U$ implement the Stinespring dilation of the quantum operation $T$. The initial state of the dilated space can be regarded to be in the pure state $\hat\rho_\mathrm{env}=\ket e\bra e$, since a mixed state can be purified. The purification of $T^*(\rho)$ is then given by $U\ket{\psi}\ket{e}$ and of $T^*(\sigma)$ by $U\ket{\phi}\ket{e}$. By Uhlmann's theorem
\begin{align}
F(T^*(\hat\rho),T^*(\hat\sigma))&\geq |\braket{\psi}{\braket{e}{U^\dagger U|e}|\phi}|^2\nonumber\\
&=|\braket{\psi}{\phi}|^2\nonumber\\
&=F(\hat\rho,\hat\sigma).
\end{align}\begin{flushright}$\Box$\end{flushright}

In conclusion a theorem, that I will need in chapter~\ref{chap:trade}.
\begin{theorem}[Strong Concavity And Joint-Concavity]
Let $\hat\rho=\sum_i p_i\hat\rho_i$ and $\hat\sigma=\sum_i q_i \hat\sigma_i$ be two mixed states. Then
\begin{equation}
F(\hat\rho,\hat\sigma)\geq \sum_i p_i q_i F(\hat\rho_i,\hat\sigma_i).
\end{equation}
This property is called \emph{strong concavity}. It directly implies \emph{joint-concavity}, i.e. if $q_i=p_i$, then
\begin{equation}
F(\hat\rho,\hat\sigma)\geq \sum_i p_i^2 F(\hat\rho_i,\hat\sigma_i).
\end{equation}
\end{theorem}
\emph{Proof:} Suppose $\ket{\psi_i}$ and $\ket{\phi_i}$ are the purifications of $\hat\rho_i$ and $\hat\sigma_i$. Then $\ket\psi\equiv\sum_i \sqrt{p_i}\ket{\psi_i}\ket{i}$ and $\ket\phi\equiv\sum_i \sqrt{q_i}\ket{\phi_i}\ket{i}$ are the purifications of $\hat\rho$ and $\hat\sigma$ with $\ket{i}$ pure states on an ancillary system. Strong concavity and consequently joint-concavity follow directly from Uhlmann's theorem:
\begin{align}
F(\hat\rho,\hat\sigma)&\geq |\braket{\psi}{\phi}|^2\nonumber\\
&=\sum_i p_i q_i F(\hat\rho_i,\hat\sigma_i).
\end{align}\begin{flushright}$\Box$\end{flushright}

\subsubsection{Fidelity Of Quantum Operations}
In order to judge the quality of a quantum operation, it is necessary to compare an input state with the output state of the operation. In fact, an appropriate figure of merit judges how close an operation is to identity. Because the quality is determined by the state that is influenced the most by the operation, worst case performance must be consideren in the definition of the fidelity of a quantum operation. 

\begin{definition}[Fidelity Of A Quantum Operation]
Let $T^* : M_d \to M_d$ be a quantum operation and $T:M_d \to M_d$ its dual. The fidelity of the quantum operation is
\begin{equation}
F_\mathrm{wc}(T)=\mathrm{inf}_{\hat \rho~\mathrm{pure}~}F(\hat\rho,T(\hat\rho)).
\end{equation}
\end{definition}

In this definition only pure states are included, since, because of joint-concavity of fidelity, the infimum is found among pure states. It reflects the fact that mixed states are always less (or equally) distorted by a quantum operation than pure states.

\newpage

\chapter{Quantum Measurements}
\label{chap:qmeas}

In this chapter, I will discuss measurements and measurement instruments (section~\ref{sec:qmeas}). Thereafter, a special class of instruments, so-called covariant instruments, is classified in section~\ref{sec:cm}. This classification plays an important role in chapter~\ref{chap:trade}.

\section{Measurement Instruments}
\label{sec:qmeas}

Before treating measurement instruments, I will start with common quantum measurement.

\subsection{Introduction To Quantum Measurement}
In some sense, it is not hard to define quantum measurement: the processing of quantum information to classical information. Mathematical formulation though is a little harder. 

\subsubsection{Classical Quantum Measurement}
Introductory courses to quantum mechanics define quantum measurement by a set of measurement operators $\{M_i\}$, each of which corresponds to an outcome, labeled by a subscript $i$. The probability to measure outcome $i$, if before measurement, the system is in pure state~\footnote{I take a pure state just for simplicity.} $\hat \rho = \ket{\psi}\bra{\psi}$, is
\begin{equation}
p_i=\braket{\psi}{M_i^\dagger M_i|\psi}.
\end{equation}

Normalization of the probabilities $\sum_i p_i=1$ demands 
\begin{equation}
\sum_i M_i^\dagger M_i=\mathbb{I}.
\end{equation}

If the state after measurement is not of interest and thus can be disregarded, the operators $\{Q_i=M_i^\dagger M_i\}$ suffice to describe the measurement procedure. Such operators, or the map $i \mapsto Q_i$, is called \emph{Positive Operator-Valued Measure} (POVM). 

Besides a classical outcome $i$, measurement yields a conditional state, i.e. the state after measurement. This is given by
\begin{equation}
\hat\rho_i=\frac{M_i^\dagger \hat\rho M_i}{\mathrm{Tr}(M_i^\dagger \hat\rho M_i)}.
\end{equation}
If the outcome of the measurement is unknown, the (averaged) output state is the sum over all conditional states weighted with the probability that they occur, i.e.
\begin{equation}
\hat\rho_f=\sum_i \mathrm{Tr}(M_i^\dagger \hat\rho M_i)\frac{M_i^\dagger \hat\rho M_i}{\mathrm{Tr}(M_i^\dagger \hat\rho M_i)}=\sum_i M_i^\dagger \hat\rho M_i.
\end{equation}

\subsubsection{POVMs}
A measurement result is given by a choice out of some set of measurement outcomes. Let $\Omega$ be such a measurable set of outcomes of a measurement procedure (the labels $i$ from above). Then $\Sigma(\Omega)$ is the $\sigma$-algebra over the set $\Omega$. If $\Omega$ is a finite set, then $\Sigma(\Omega)$ is the set of all subsets of $\Omega$. In general, a quantum measurement is an affine map $\hat\rho \to \mu_{\hat\rho}(\D\omega)$ with $\omega\in\Omega$ of $S$ into the set of all probability distributions on $\Omega$. 

There is a one-to-one correspondence between measures $\mu_{\hat\rho}(\D\omega)$ and a so called resolution of identity $Q = \{Q(A); A\in \Sigma(\Omega)\}$. The elements of the set $Q$ satisfy for all $A\in\Sigma(\Omega)$
\begin{enumerate}
	\item $Q(\Omega)=\mathbb{I}$;
	\item $Q(A)\geq 0$;
	\item $Q(\bigcup_{i=1}^\infty A_i)=\sum_{i=1}^\infty Q(A_i)$ for disjoint $A_i$.
\end{enumerate}
The correspondence between $\mu_{\hat\rho}(\D\omega)$ and $Q$ is given by
\begin{equation}
 \mu_{\hat\rho}(A)=\mathrm{Tr}(\hat\rho Q(A)).
\end{equation}
This expression is interpreted as the probability to measure an outcome in $A$ when the system is in state $\hat\rho$. 

As described above, in simple quantum mechanics, measurement is given by a set of operators each of which belongs to a particular outcome of the measurement. Now, suppose $\Omega$ is a finite set and $\{Q(u)=Q_u;u\in \Omega\}$ is a set of Hermitian operators (observables) that satisfy $\sum_{u\in\Omega} Q_u=\mathbb{I}$ and $Q_u\geq 0$, then
\begin{equation}
Q(A) =\sum_{u\in A} Q_u,\qquad A\subset \Omega
\end{equation}
recovers this aspect of measurement. If in addition
\begin{equation}
Q(A_1)Q(A_2)=0,\qquad\mathrm{if}~A_1\cap A_2=\emptyset,
\end{equation}
i.e. the resolution of identity is orthogonal, then a projection-valued Von Neumann measurement is defined.

\subsection{Quantum Instruments}

\begin{figure}[!htb]
\begin{center}
\includegraphics[width=8cm]{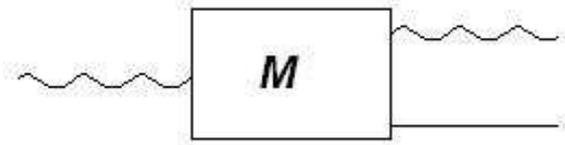}
\caption{An instrument. }
\label{fig:M}
\end{center}
\end{figure}

Measurement provides us with a classical outcome and a conditional state. Fig.~(\ref{fig:M}) is an illustration of a measurement device $M$. I will call it a measurement instrument~\cite{Dav}, or shorter, an instrument. In general, an instrument is a quantum operation, defined in the Heisenberg picture by
\begin{equation}
M : \cal A \otimes \cal B \to \cal A,
\end{equation}
where $\cal A$ is the algebra of observables of the system and the Abelian algebra $\cal B$ captures the classical outcomes of the measurement instrument. If the dimension of the system is finite, then $\cal A=\B$ and ${\cal B}={\cal L}^\infty(\Omega)$, the functions $f: \Omega \to \mathbb{C}$ with $\Omega$ a finite set. So the wavy lines in fig~(\ref{fig:M}) correspond to quantum information and the straight line to classical information. 

\begin{figure}[!htb]
\begin{center}
\includegraphics[width=8cm]{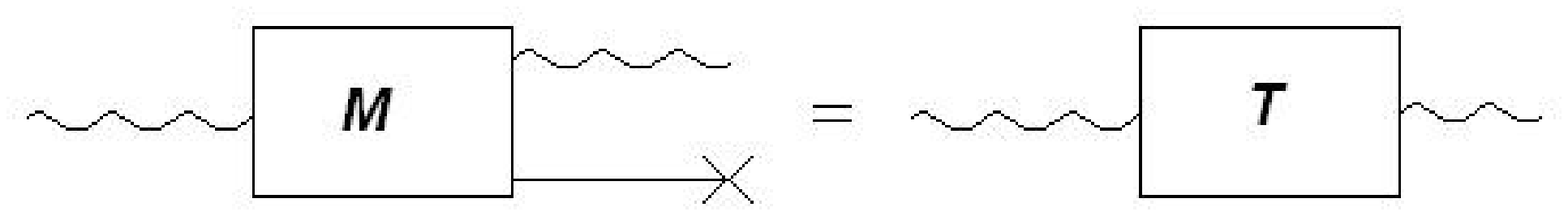}
\caption{The through-going part of an instrument.}
\label{fig:T}
\end{center}
\end{figure}

The through-going channel $T$ of an instrument covers the change of the initial state. It is a quantum operation
\begin{equation}
T : \cal A \to \cal A
\end{equation}
and is obtained by disregarding (or throwing-away) the classical outcome of the instrument. Since throwing-away measurement outcomes is equal to weigthed averaging over all possible outcomes, $T$ covers the averaged state after measurement. See fig.~(\ref{fig:T}) for an illustration. 

\begin{figure}[!htb]
\begin{center}
\includegraphics[width=8cm]{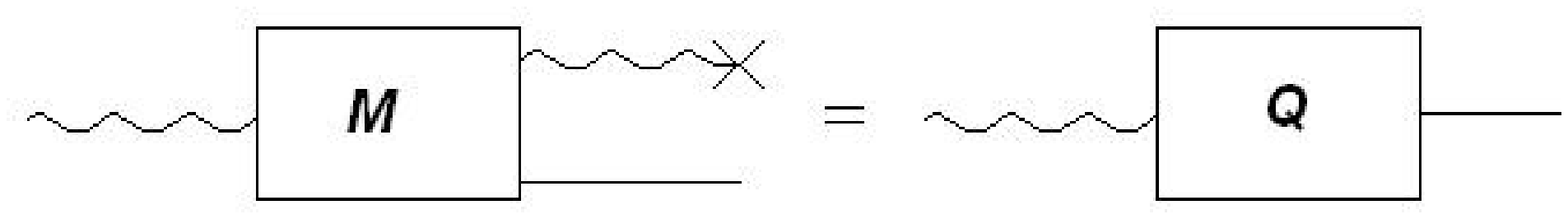}
\caption{The POVM part of an instrument.}
\label{fig:Q}
\end{center}
\end{figure}

The measurement channel that provides the outcome of the measurement procedure is the POVM
\begin{equation}
Q : \cal B \to \cal A.
\end{equation}
It is obtained by disregarding the conditional state, i.e. tracing over the conditional state. See fig.~(\ref{fig:Q}). For finite dimension, $Q : {\cal C}(\Omega) \to M_d$ and a POVM $\tilde Q:\Sigma(\Omega) \to M_d$ as defined above are related via
\begin{equation}
Q(f) = \int_{\Omega} f(\omega)\tilde Q(\D\omega),\qquad \omega\in\Omega.
\end{equation}
The function $f$ is a function on the outcomes. I will use both equivalent definitions of a POVM. It will follow from the setting and argument of $Q$ which one is used. 

It is important to stress that by the Stinespring dilation theorem an arbitrary instrument is given by
\begin{equation}
M(a,f)=V^*a\otimes P(f)V
\end{equation}
with $P(f)$ a POVM on the ancillary space. The fact that the POVM is defined on the ancillary space is a consequence of corollary 3.2 of Ref~\cite{Rag}. 

\begin{corollary}[\cite{Rag}]
\label{cor:radpovm}
Let $S, T : {\cal B(H}_1) \to {\cal B(H}_2)$ be two CP maps. And let $T(a)=V^*a\otimes \mathbb{I}V$ be the Stinespring dilation of $T$. Then $S\leq T$ if and only if there exists a positive operator $0 \leq F \leq \mathbb{I}$, such that
\begin{equation}
S(a)=V^*a\otimes F V
\end{equation}
\end{corollary}

In this corollary $S\leq T$ means that $||S(a)\psi||\leq ||T(a)\psi||$ for all $a\in{\cal B(H}_1)$ and all $\psi\in{\cal H}_2$. The following corollary follows immediately.

\begin{corollary}
\label{cor:instrpovm}
Let $M:\B\times\Sigma(\Omega)\to\B$ be an instrument. Let $T(a)\equiv M(a,\Omega)$ be the through going channel with Stinespring dilation $T(a)=V^* a\otimes \mathbb{I} V$. Since $M(a,A)\leq T$,
\begin{equation}
M(a,A)=V^*a\otimes F_A V.
\end{equation}
with $F_A$ a positive operator satisfying $0\leq F_A \leq \mathbb{I}$. So $F_A$ is a POVM on the ancilla space of the Stinespring dilation. 
\end{corollary}

Corollary~\ref{cor:instrpovm} implies that the POVM $Q(F)$ defined by $Q(f)=M(\mathbb{I},f)$ is given by $Q(f)=V^*\mathbb{I}\otimes P(f)V$ with $P(f)$ a POVM on the ancilla space.

\begin{example}[Measurement Of A Qubit]
Consider a two-dimensional quantum system, or in other words a qubit. The algebra of observables of this finite-dimensional system is the algebra of all complex $2\times 2$-matrices $M_2$. Suppose we want to measure this qubit in the computational basis ($\sigma_z$). An instrument that implements this measurement device, is defined by two operators $P_0=\ket 0 \bra 0$ and $P_1=\ket 1 \bra 1$. 

The measurement outcomes ($+1$ and $-1$) are provided by the POVM $Q$:
\begin{eqnarray}
Q(+1)\equiv P_0^2=\ket 0 \bra 0;\\
Q(-1)\equiv P_1^2=\ket 1 \bra 1,
\end{eqnarray}
The probability to measure $+1$ is $\mathrm{Tr}(\hat\rho P_0)=\mathrm{Tr}(\hat\rho Q(+1))$ and to measure $-1$ is $\mathrm{Tr}(\hat\rho P_1)=\mathrm{Tr}(\hat\rho Q(-1))$.

The corresponding to the conditional states is
\begin{eqnarray}
\hat\rho_{+1}= \frac{P_0\hat\rho P_0}{\mathrm{Tr}(\hat\rho P_0)};\\
\hat\rho_{-1}= \frac{P_1\hat\rho P_1}{\mathrm{Tr}(\hat\rho P_1))}.
\end{eqnarray}

The through-going channel is the averaged state after measurement, which is in the Sch\"odinger picture
\begin{align}
T^*(\hat \rho)&\equiv P_0\hat\rho P_0 + P_1\hat\rho P_1 \nonumber\\
&= \ket 0 \bra 0\hat\rho \ket 0 \bra 0 + \ket 1 \bra 1\hat\rho \ket 1 \bra 1.
\end{align}
\end{example}

\begin{example}[Rotating Polarizer Measurement]
\label{ex:rotpol}
Suppose we don't want to measure the qubit in a fixed basis, but in a randomly chosen basis. The instrument implementing this measurement is defined by all one-dimensional projection operators $p=\ket\psi\bra\psi$ (i.e. pure states). The POVM $Q$ is defined over the space of all one-dimensional projections $\PO$ and is given by
\begin{equation}
Q(f)=2\int_\PO p f(p) \D p
\end{equation}
with $f(p)$ a function depending on pure state $p$ and the factor $2$ for normalization. It is an infinite set of operators as opposed to the finite set of example I. The through-going channel is defined by
\begin{equation}
T^*(\hat\rho)=2\int_\PO p \hat\rho p \D p.
\end{equation}

The POVM and through-going channel together are obtained from an instrument $M$, which in this case is given by
\begin{equation}
M(a,f)=2\int_\PO p a p f(p) \D p.
\end{equation}
The two outputs are then defined by $Q(f)=M(\mathbb{I},f)$ and $T(a)=M(a,1)$. 

Observe in addition that
\begin{align}
\mathrm{Tr}(T^*(\hat\rho)a)&=\mathrm{Tr}(\hat\rho T(a))\nonumber\\
&=\int_\PO \mathrm{Tr}(\hat\rho pap) \D p\nonumber\\
&=\int_\PO \mathrm{Tr}(p\hat\rho pa) \D p
\end{align}
and consequently $T^*(\hat\rho)=T(\hat\rho)$. 
\end{example}

A qubit may be any two-dimensional system. Two examples are the spin of spin-$1/2$ systems and the polarization of a photon. The latter may illustrate example~\ref{ex:rotpol}. Measurement of the polarization as described above is equivalent to measurement in a randomly chosen basis. This is realized by placing a rotating polarizer before a measurement device that measures the polarization of the photon in a fixed basis. 

Measurement of this kind is called covariant measurement. More on covariant instruments is found in section~\ref{sec:cm}. In fact, it turns out that optimal measurement is covariant. Optimal measurement provides the best trade-off between the quality of the outcome and distortion of the input state. See chapter~\ref{chap:trade}.

\subsection{Heisenberg Uncertainty Relations}
The Heisenberg Uncertainty Principle is a directly related to the Heisenberg Uncertainty Relations. 

\begin{theorem}[Heisenberg Uncertainty Relations]
Let $a:\Hi\to \Hi$ and $b:\Hi\to \Hi$ be two Hermitian observables or measurement operators. Let the variance $\Delta(x)$ of an operator $x$ be defined by
\begin{equation}
(\Delta(x))^2=\expec{(x-\expec{x})^2}=\expec{x^2}-\expec{x}^2.
\end{equation}
Then
\begin{equation}
\Delta(a)\Delta(b)\geq\frac{1}{2}|\expec{[a,b]}|.
\end{equation}
\end{theorem}
\emph{Proof:} Since for Hermitian operators
\begin{equation}
\frac{1}{4}|\expec{ab-ba}|^2 \leq \frac{1}{4}|\expec{ab-ba}|^2 + \frac{1}{4}|\expec{ab+ba}|^2 = |\expec{ab}|^2 
\end{equation}
and 
\begin{equation}
|\expec{ab}|^2 \leq \expec{a^2}\expec{b^2}
\end{equation}
by the Cauchy-Schwarz inequality, the Robertson-Schr\"odinger equation holds
\begin{equation}
\frac{1}{4}|\expec{[a,b]}|^2 \leq \expec{a^2}\expec{b^2}.
\end{equation}
The theorem is now readily proved by substitution, $a\to a - \expec a$ and $b\to b - \expec b$.\begin{flushright}$\Box$\end{flushright}

\newpage
\section{Covariant Measurement}
\label{sec:cm}

Covariant measurement plays a crucial role in the proof of the optimal trade-off theorem to be presented in chapter~\ref{chap:trade}. As a matter of fact, many optimal devices, such as the optimal spin-flipping device~\cite{Wer3, Gis} and the optimal pure state quantum cloner~\cite{Wer2}, are covariant instruments. 

This section will contain the classification of the family of all covariant instruments, quantum operations and POVMs. 

\subsection{Covariance}

Let $G $ be a locally compact group, carrying a Haar measure $\mu(dg)$, which allows to integrate functions defined on $G$. Let $\Omega$ be a locally compact space, which is called a $G$-space if there exists a jointly continuous map $L: G \times \Omega \to \Omega$, called the action of $G$ on $\Omega$, such that
\begin{equation}
g_1 (g_2 x) = (g_1 g_2)x
\end{equation}
for all $g_1,g_2 \in G$ and $x\in \Omega$. The map $L$ is called transitive if for some $x_0 \in \Omega$ and for all $x\in \Omega$ there exists $g\in G$ such that $gx_0 = x$. Transitivity of a $G$-space means that every element of the space is reached from any other element via the map $L$.

The stability subgroup $H_{x_0}$ of a point $x_0 \in \Omega$ is defined by
\begin{equation}
H_{x_0} = \{g\in G | gx_0 = x_0\}.
\end{equation}
This group $H_{x_0}$ is a closed subgroup of $G$. For the transitive $G$-space $\Omega$ and the stability group $H$ of some element $x_0$, there is a one-to-one correspondence between the right cosets $Hg$ of $H$ and the points $gx$ of $\Omega$. If $\Omega$ and $G$ are second countable, i.e. the topology has a countable base, this correspondence is a homeomorphism and a $G$-space isomorphism of $\Omega$ with the set $G/H$ of left cosets of $H$, given its quotient topology and the induced action of $G$, i.e. $\Omega \simeq G/H$. Now, if $H$ is compact, there exists a Haar measure $\nu(dx)$ on $\Omega$, such that
\begin{displaymath}
\xymatrix{
G \ar[d]^{\rho,\qquad \nu(dx)=\mu(\rho^{-1}(dx))}\\
\Omega\simeq G/H}
\end{displaymath}

I will restrict to finite dimension, so the Hilbert space describing the system is ${\cal H}=\C$. The symmetry group of density operators on this system of finite dimension $d$ is $U(d)$, the Lie group of unitary matrices. Since $U(d)$ only differs a phase factor from $SU(d)$, the Lie group of complex $d\times d$-matrices with unit determinant, the analysis of covariant instruments can be restricted to $SU(d)$-covariance. 

The representation $U: SU(d)\to \B$ of $SU(d)$ on $\C$ is denoted by $u_g:=U(g)$ in which the subscript $g$ will be dropped unless necessary. 

\subsubsection{Covariant Instruments, Quantum Operations and POVMs}

A quantum operation, or a CP map in general, $T : \B \to \B$, is called covariant if it satisfies 
\begin{equation}
T(u_gau_g^*)=u_gT(a)u_g^*,\qquad a \in \B, \forall g\in SU(d).
\end{equation}
This means that the output state of a covariant quantum operation rotates along with rotation of the input state. 

A POVM $Q : \Sigma(\Omega) \to \B$ is called covariant if
\begin{equation}
Q(g^{-1}A)=u_gQ(A)u_g^*,\qquad A\in \Sigma(\Omega).
\end{equation}
This implies that the measure on the outcome, which $Q(A)$ actually expresses, transforms along with the rotation of the input state.

Covariance of an instrument $M: \B \times \Sigma(\Omega) \to \B$ is expressed by
\begin{equation}
M(u_gau_g^*,g^{-1}A)=u_gM(a, A)u_g^*,\qquad A\in \Sigma(\Omega), a\in \B.
\end{equation}

\begin{example}[Rotating Polarizer Measurement]
The POVM defined in example~\ref{ex:rotpol} is clearly covariant:

\begin{equation}
u_g^*Q(g^{-1}f)u_g=2\int_\PO u_g^*pu_g f(u_g^*pu_g) \D(u_g^*pu_g)=2\int_\PO p f(p)\D p=Q(f)
\end{equation}
since $\D p$ is unimodular (both left- and right-invariant under the action of $SU(d)$) and $\PO$ is transitive, i.e. $g^{-1}\PO=\PO$. Because the through-going channel is also covariant,
\begin{equation}
u_g^*T^*(u_g\hat\rho u_g^*)u_g=2\int_\PO u_g^*p u_g \hat\rho u_g^* pu_g \D p = 2\int_\PO p \hat\rho p \D p = T^*(\hat\rho),
\end{equation}
the instrument $M$ is covariant as well:
\begin{align}
u_g^*M(u_gau_g^*, g^{-1}f)u_g &=2\int_\PO u_g^*pu_g a u_g^*pu_g f(u_g^*pu_g) \D(u_g^*pu_g)\nonumber\\
&=2\int_\PO p a p f(p) \D p\nonumber\\
&=M(a, f).
\end{align}
\end{example}

\subsubsection{Classification Of Covariant Instruments}
The classification of $SU(d)$-covariant instruments on finit-dimensional systems will be carried out in three steps. First I will classify the family of covariant quantum operations. Thereafter I will classify the covariant POVMs. At the end, these two families are combined to form the family of covariant instruments. 

\subsection{Covariant Quantum Operations}
\label{sec:comap}
Covariant CP maps have the following simple characterization.
Let $T : \B \to \B$ be a covariant CP map. Then it is given by
\begin{equation}
T(a) = \left(1 - \alpha\frac{d^2}{d^2-1}\right)a + 
\left(\alpha\frac{d}{d^2-1}\mathrm{Tr}(a)\right)\mathbb{I}\qquad 0\leq \alpha \leq 1
\end{equation}
with $d$ the dimension of $\Hi$ and $\alpha \in \mathbb{R}$. Covariance of the operation $T(a)$ means that it intertwines the trivial and adjoint representation ($g\mapsto u_g \dot u_g^*$). So $T(a)$ is easily found, since it must commute with all unitaries that commute with $a$. The restriction of the factor $\alpha$ is because of complete-positivity. 

A general CP map $T : \B \to \B$ is, by the Stinespring dilation theorem, given by
\begin{equation}
T(a)=V^*a\otimes\mathbb{I}_{{\cal E}}V
\end{equation}
with $V:\C \to \C \otimes{\cal E}$. In this form, the covariance property of CP map is
\begin{equation}
uV^*a\otimes\mathbb{I}_{{\cal E}}Vu^*=V^*uau^*\otimes\mathbb{I}_{{\cal E}}V.
\end{equation}

In order to find all covariant CP maps, all operators $V$ that satisfy this equation have to be characterized. Define $D_g : \C\otimes{\cal E} \to \C\otimes{\cal E}$ by
\begin{equation}
D_g: a \otimes \mathbb{I}_{{\cal E}}V\psi \mapsto au_g\otimes \mathbb{I}_{{\cal E}}Vu_g^*\psi\qquad \psi\in\Hi.
\end{equation}
The subscript $g$ may be omitted unless necessary. 

\begin{lemma}
The operator $D$ extends to a unitary representation of $SU(d)$ on $\cal E$.
\end{lemma}
\textit{Proof:} Because the operator $D$ commutes with all $a\otimes \mathbb{I}_{{\cal E}}$,
\begin{align}
&[D,a\otimes\mathbb{I}_{{\cal E}}](\mathbb{I} \otimes \mathbb{I}_{{\cal E}}V\psi)\nonumber\\
& = (D(a\otimes \mathbb{I}_{{\cal E}})- (a\otimes \mathbb{I}_{{\cal E}})D)(\mathbb{I}\otimes \mathbb{I}_{{\cal E}} V\psi)\nonumber\\
& = D(a\otimes\mathbb{I}_{{\cal E}}V\psi) - (a\otimes \mathbb{I}_{{\cal E}})(u^*\otimes \mathbb{I}_{{\cal E}}Vu\psi)\nonumber\\
& = (au^*\otimes \mathbb{I}_{{\cal E}}Vu\psi) - (au^*\otimes \mathbb{I}_{{\cal E}}Vu\psi)\nonumber\\
& = 0,
\end{align}
it is an operator on ${\cal E}$. It is a representation on $\cal E$, while
\begin{align}
D_{g_1}D_{g_2}(a \otimes \mathbb{I}_{{\cal E}}V\psi)&=au_{g_1}u_{g_2}\otimes \mathbb{I}_{{\cal E}}u_{g_2}^*u_{g_1}^*V\psi\nonumber\\
&=au_{g_1}u_{g_2}\otimes \mathbb{I}_{{\cal E}}(u_{g_1}u_{g_2})^*V\psi\nonumber\\
&=D_{g_1 g_2}(a \otimes \mathbb{I}_{{\cal E}}V\psi).
\end{align}
In conclusion, $D$ is unitary:
\begin{align}
& \braket{D(a\otimes\mathbb{I}_{{\cal E}}V\psi)}{D(b\otimes\mathbb{I}_{{\cal E}}V\phi)}\nonumber\\ 
& = \braket{au\otimes\mathbb{I}_{{\cal E}}Vu^*\psi}{bu\otimes\mathbb{I}_{{\cal E}}Vu^*\phi}\nonumber\\
& =\braket{\psi}{uV^*u^*a^*bu\otimes\mathbb{I}_{{\cal E}}Vu^*\phi}\nonumber\\
& =\braket{\psi}{V^*a^*b\otimes\mathbb{I}_{{\cal E}}V\phi}\nonumber\\
& =\braket{a\otimes\mathbb{I}_{{\cal E}}V\psi}{b\otimes\mathbb{I}_{{\cal E}}V\phi}\nonumber
\end{align}
in which I used the covariance property in the third line. So consequently, the operator $D$ is a unitary representation of $SU(d)$ on the ancilla Hilbert space ${\cal E}$. \begin{flushright}$\Box$\end{flushright}

\begin{corollary}
The Stinespring dilation operators $V$ of covariant CP maps intertwine $u$ and $u\otimes D$:
\begin{equation}
u\otimes DV=Vu.
\end{equation}
\end{corollary}

\subsubsection{Decomposing The Tensor Representation}

The dilation operator $V$ intertwines $u$ and $u\otimes D$. By Schur's lemma and simple reducibility of $SU(d)$, such operators are non-zero if and only if $u$ is contained at least once in the decomposition of $u\otimes D$. 

Representations of $SU(d)$ are decomposed in irreducible representations by the Clebsch-Gordan formula. I will use Young diagrams to find the Clebsch-Gordan decompostion of $D$~\cite{You}. It follows from the analysis of Young tableaux, that the tensor product $u\otimes D$ contains $u$ only if one of the irreducible representations of $D$ is either the trivial representation $\mathrm{Triv}$ or the adjoint representation $\mathrm{Ad}$, the representation of a Lie group on itself via conjugation. 

The Young diagram of the defining representation $u$ on $\C$ is given by

\setlength{\unitlength}{0.5cm}
\begin{picture}(2,1.5)
\put(0,0){\line(0,1){1}}
\put(0,0){\line(1,0){1}}
\put(0,1){\line(1,0){1}}
\put(1,0){\line(0,1){1}}
\put(1.3,0.3){$(u).$}
\end{picture}

Using the rules for the tensor product of two Young diagrams (see appendix~\ref{sec:young}), it is not hard to see that the dot in the diagram

\setlength{\unitlength}{0.5cm}
\begin{picture}(8,1.5)
\put(0,0){\line(0,1){1}}
\put(0,0){\line(1,0){1}}
\put(0,1){\line(1,0){1}}
\put(1,0){\line(0,1){1}}
\put(1.3,0.3){$(u)$}
\put(2.3,0.3){$\otimes$}
\put(3.3,0.5){\circle*{0.4}}
\put(3.8,0.3){$=$}
\put(4.8,0){\line(0,1){1}}
\put(4.8,0){\line(1,0){1}}
\put(4.8,1){\line(1,0){1}}
\put(5.8,0){\line(0,1){1}}
\put(6.1,0.3){$(u)$}
\put(7.3,0.3){$\oplus$}
\put(8.0,0.3){$\ldots$}
\end{picture}

must be

\setlength{\unitlength}{0.5cm}
\begin{picture}(6,4)
\put(0.5,1.5){\circle*{0.2}}
\put(1,1.3){(Triv)}
\put(3.2,1.3){or}
\put(4.3,0){\line(1,0){1}}
\put(4.3,0){\line(0,1){2.1}}
\put(4.3,1){\line(1,0){1}}
\put(5.3,0){\line(0,1){2.1}}
\put(4.5,0.4){\tiny{d-1}}
\put(4.8,1.3){\circle*{0.1}}
\put(4.8,1.55){\circle*{0.1}}
\put(4.8,1.8){\circle*{0.1}}
\put(4.3,2.1){\line(1,0){2}}
\put(4.3,2.1){\line(0,1){1}}
\put(4.3,3.1){\line(1,0){2}}
\put(5.3,2.1){\line(0,1){1}}
\put(6.3,2.1){\line(0,1){1}}
\put(4.7,2.5){\tiny{1}}
\put(6.6,1.3){(Ad)}
\end{picture}

The dimension of the trivial representation Triv is $1$ and the dimension of the adjoint representation Ad is $d^2-1$. The fact that the number of dilation operators is always less than $d^2$ and thus the dimension of $\cal E$ is always less than $d^2$, implies that if $D$ contains only trivial representations, all CP maps are trivial. If $D$ just contains an adjoint representation, there will be only one CP map. The minimal dilation allows to define ${\cal E}=\C\otimes\C$. 

On this space $\mathrm{Triv}\oplus\mathrm{Ad}$ is unitarily equivalent to the tensor representation $\bar{u}\otimes u$. Here the representation $\bar{u}_{ij}$ is defined by $\bar{u}_{ij}=\overline{u_{ij}}$. 

Finding all covariant quantum operations, comes down to finding the decomposition of $u\otimes\bar{u}\otimes u \cong u \otimes( \mathrm{Triv}\oplus\mathrm{Ad})$. Note that by the rules for Young tableaux, 
\begin{align}
u\otimes\bar{u}\otimes u & \cong u\otimes\mathrm{Triv}\oplus\mathrm{Ad}\nonumber\\
& = u\oplus (u\otimes \mathrm{Ad})\nonumber\\
& = u \oplus u \oplus \ldots
\end{align}
In this equation I used that $u\otimes\mathrm{Ad}$ contains, amongst others, $u$. 

\subsubsection{The Covariant Stinespring Dilation}
The dilation operators $V$ must couple a vector $\psi\in\C$ to vectors in $\C\otimes\C$ that are invariant under the action of $\bar u \otimes u$. Let ${\cal E}_1$ and ${\cal E}_2$ be two subspaces of $\Hi \otimes {\cal E} = \C \otimes \C \otimes \C$ that are formed by such vectors. Let $\psi_0$ be defined by
\begin{equation}
\psi_0 := \frac{\sum_{i=1}^d e_i\otimes e_i}{\sqrt{d}}.
\end{equation}
in which $\{e_i\}$ form an orthonormal basis in $\C$. It is clear that 
\begin{eqnarray}
(u\otimes\bar{u}\otimes u) \psi\otimes\psi_0 =u\psi\otimes\psi_0,\\
(u\otimes\bar{u}\otimes u) \psi_0\otimes\psi =\psi_0\otimes u\psi.
\end{eqnarray}

As exemplification of this, consider $d=2$. Then
\begin{align}
(u\otimes\bar{u}\otimes u) \psi\otimes\psi_0 &= u\psi\otimes\left(
\left(\begin{array}{c c}
\bar{a} & \bar{b} \\
-b & a
\end{array}\right)\otimes
\left(\begin{array}{c c}
a & b \\
-\bar{b} & \bar{a}
\end{array}\right)\right)
\left(\begin{array}{c}
1 \\
0\\
0\\
1\end{array}\right)\nonumber\\
&=u\psi\otimes\left(\begin{array}{c}
|a|^2 + |b|^2 \\
0\\
0\\
|a|^2 + |b|^2\end{array}\right)=u\psi\otimes\psi_0
\end{align}
and similarly for $(u\otimes\bar{u}\otimes u) \psi_0\otimes\psi$. 

For arbitrary dimension, define ${\cal E}_1$ by
\begin{equation}
{\cal E}_1:=\{\psi \otimes \psi_0, \psi\in\C\}\\
\end{equation}
and ${\cal E}_2$ by
\begin{equation}
{\cal E}_2:=\{\psi \otimes \psi_0 - 2 \psi_0 \otimes \psi,\psi\in\C\}.
\end{equation}
Then ${\cal E}_1$ and ${\cal E}_2$ are orthogonal:
\begin{align}
& \braket{\phi\otimes \frac{\sum_{i=1}^d e_i\otimes e_i}{\sqrt{d}}}{\frac{1}{\sqrt{d^2-1}}\{\psi \otimes \frac{\sum_{j=1}^d e_j\otimes e_j}{\sqrt{d}} - d \frac{\sum_{j=1}^d e_j\otimes e_j}{\sqrt{d}} \otimes \psi \}}\nonumber\\
& =  \frac{1}{\sqrt{d^2-1}}\braket{\phi}{\psi}\braket{\frac{\sum_{i=1}^d e_i\otimes e_i}{\sqrt{d}}}{\frac{\sum_{j=1}^d e_j\otimes e_j}{\sqrt{d}}}\nonumber\\
&\qquad - \frac{1}{\sqrt{d^2-1}}\frac{d\sum_{i,j}\braket{\phi}{e_j}\braket{e_i}{e_j}\braket{e_i}{\psi}}{d}\nonumber\\
& = \frac{1}{\sqrt{d^2-1}}\left\{\braket{\phi}{\psi}-\sum_{i=1}^d\braket{\phi}{e_i}\braket{e_i}{\psi}\right\}\nonumber\\
& = 0.\nonumber
\end{align}

The dilation operators $V$ of covariant CP maps embed $\Hi$ in $\C \otimes \C \otimes \C$ such that $u\otimes D V= Vu$ and thus couple $\Hi$ to ${\cal E}_1$ and ${\cal E}_2$,
\begin{align}
\label{eq:covop}
V : \psi \mapsto & c_1\left(\psi\otimes\psi_0\right)+\nonumber\\
&\frac{c_2}{\sqrt{d^2-1}}\left(\psi \otimes \psi_0 - d\psi_0\otimes \psi\right).
\end{align}
with $c_1,c_2 \in \mathbb{C}$. The normalization condition $V^*V=\mathbb{I}$ yields
\begin{equation}
|c_1|^2 + |c_1|^2 = 1.
\end{equation}

The actual form of a covariant CP map $T(a)$ is obtained via
\begin{align}
\braket{\phi}{T(a)\psi} & = \braket{\phi}{V^*a\otimes\mathbb{I}V|\psi}\nonumber\\
& = \braket{V\phi}{a\otimes \mathbb{I}V|\psi}\nonumber\\
& = |c_1|^2\braket{\phi}{a\psi} + |c_2|^2\left(\frac{d}{d^2-1}\mathrm{Tr}(a)\braket{\phi}{\psi} - \frac{1}{d^2-1}\braket{\phi}{a\psi}\right)\nonumber\\
& \Rightarrow T(a) = \left(|c_1|^2 - \frac{|c_2|^2}{d^2-1}\right)a + 
\left(\frac{d|c_2|^2}{d^2-1}\mathrm{Tr}(a)\right)\mathbb{I}
\end{align}
and is finally given by
\begin{equation}
T(a) = \left(1 - \alpha\frac{d^2}{d^2-1}\right)a + 
\left(\alpha\frac{d}{d^2-1}\mathrm{Tr}(a)\right)\mathbb{I}\qquad 0\leq \alpha \leq 1
\end{equation}
with $\alpha=|c_2|^2 \in \mathbb{R}$. The result is a one-parameter family of covariant CP maps.

\begin{example}[Rotating Polarizer Measurement]
The through-going channel of the covariant measurent instrument of example~\ref{ex:rotpol} is given by
\begin{equation}
T(a)=2\int_\PO pap\D p.
\end{equation}
By theorem, it is given by
\begin{equation}
\tilde T(a) = \left(1 - \alpha\frac{4}{3}\right)a + 
\left(\alpha\frac{2}{3}\mathrm{Tr}(a)\right)\mathbb{I} 
\end{equation}
with $0\leq \alpha \leq 1$. Make use of 
\begin{equation}
\mathrm{Tr}(\hat\rho \tilde T(\hat\rho))=\left(1-\alpha \frac 43\right) + \alpha\frac 23=1-\alpha\frac 23
\end{equation}
for pure state $\hat\rho$ and 
\begin{equation}
\mathrm{Tr}(\hat\rho T(\hat\rho))=2\int_\PO \mathrm{Tr}(p\hat\rho)^2 \D p = \frac{2}{3}
\end{equation}
(see appendix~\ref{sec:trprop} and~\ref{sec:traceint}), to calculate $\alpha=1/2$ and $\tilde T$:
\begin{equation}
\tilde T (a)= \frac 13 a + \frac 13 \mathrm{Tr}(a)\mathbb{I}.
\end{equation}
This implies that the averaged state after covariant measurement of e.g. $\ket 0 \bra 0$ is
\begin{equation}
T^*(\ket 0\bra 0)=\frac 23 \ket 0 \bra 0 + \frac 13 \ket 1 \bra 1.
\end{equation}

Since the fidelity $F_\mathrm{wc}(T)$ is given by
\begin{equation}
F_\mathrm{wc}(T)=\mathrm{inf}_{\hat \rho~\mathrm{pure}~}F(\hat\rho,T(\hat\rho))=\mathrm{Tr}(\hat\rho T(\hat\rho)),
\end{equation}
in which inf could be disregarded because of covariance, the fidelity of the averaged state of this measurement is given by
\begin{equation}
F_\mathrm{wc}(T)=1-\alpha\frac 23.
\end{equation}
So with $\alpha=1/2$, the fidelity is $2/3$. This value is proved to correspond to optimal measurement~\cite{Bru}. The value $\alpha=0$ corresponds to no distortion, i.e. $F_\mathrm{wc}(T)=1$. The value $\alpha=1$ corresponds to the fidelity $F_\mathrm{wc}(T)=1/3$ of a universal \textsc{not-}gate. See Ref.~\cite{Wer3}. 
\end{example}

\subsection{Covariant POVMs}
Besides the one-parameter family of covariant quantum operations, there is a family of covariant POVMs. This family is classified by a theorem by Holevo.
\begin{theorem}[Holevo]
\label{the:hol}
Let $P_0$ be a Hermitean positive operator in the representation space $U(d)$, commuting with the operators $\{u_g;g\in H_{x_0}\}$ of the stability group and satisfying
\begin{equation}
\int_G u_gP_0 u_g^*\mu(dg)=\mathbb{I}
\end{equation}
Then an operator-valued function of $x_0$ defined by
\begin{equation}
\label{eq:p0}
P'(A)=\int_{gx_o\in A} u_gP_0 u_g^*\mu(dg)\equiv\int_A P(\omega)\D\omega
\end{equation}
is a POVM, covariant with respect to $g\mapsto u$. Conversely, for any covariant POVM $P'(A)$, there is a unique Hermitean positive operator $P_0$, satisfying eq.~\ref{eq:p0} such that $P'(A)$ is expressed by the above construction.
\end{theorem}

\emph{Proof:} The proof of this theorem is found in Ref~\cite{Hol}. The first statement follows easily by checking positivity, $\sigma$-additivity and normalization of $P'(A)$. 

The crucial element in the proof of the converse statement is a Radon-Nikodym like theorem, which proves that for any covariant measurement $P'(A)$, there exists a unique \emph{operator density} $P(\omega)$, such that
\begin{equation}
P'(A)=\int_A P(\omega)\D\omega.
\end{equation}
Covariance implies
\begin{equation}
\int_A u_g^* P(\omega)u_g\D\omega = \int_{g^{-1}A} P(\omega)\D\omega = \int_A P(g^{-1}\omega)\D\omega
\end{equation}
so that by uniqueness of the operator density
\begin{equation}
u_g^*P(\omega)u_g=P(g^{-1}\omega).
\end{equation}
Define $P_0=P(\omega_0)$ so that $P(\omega)=u_gP_0u_g^*$. This closes the proof.\begin{flushright}$\Box$\end{flushright}

\subsubsection{Covariant Measurement On $\C$}
\begin{figure}[!htb]
\begin{center}
\includegraphics[width=4cm]{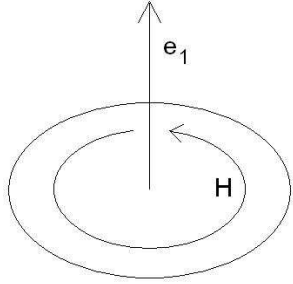}
\caption{The only two operators commuting with rotation of the plane orthogonal to $e_1$ are projection on $e_1$ and projection on the orthogonal plane.}
\label{fig:cova}
\end{center}
\end{figure}

The stability group $H$ of some pure state $\hat\rho=\ket{e_1}\bra{e_1}$ is a subgroup of $SU(d)$, namely the circle group $U(d-1)$. The positive, Hermitean operator $P_0$ has to commute with the representation of this group. Because the representation space $U(H)$ of $H\simeq U(d-1)$ is $\C\ominus \mathbb{C}e_1$, the commutant is given by (see fig.~(\ref{fig:cova})),
\begin{equation}
U(H)'=\{P_{e_1},P^\perp_{e_1}\}.
\end{equation}

The operator $P_0$, called the seed of the POVM, is a linear combination of these two operators;
\begin{equation}
P_0=\lambda P_{e_1}+\mu P^\perp_{e_1}.
\end{equation}
The normilzation condition $\int_G uP_0 u^*\mu(dg)=\mathbb{I}$ is used to obtain
\begin{equation}
P_0=\gamma d P_{e_1}+(1-\gamma)\frac{d}{d-1} P^\perp_{e_1}
\end{equation}
with $\gamma:=\lambda/d$. The range of $\gamma$ is restricted to $0\leq\gamma\leq 1$ because of positivity of $P_0$.

\subsection{Covariant Instruments}
\label{sec:covinstr}
The most general form of an instrument $M(a,A)$ is given by 
\begin{equation}
M(a,A)=V^*a\otimes P(A)V,\qquad A\in \Sigma(\Omega),  a\in \B,
\end{equation}
where $P : \Sigma(\Omega) \to {\cal B(E)}$ is a POVM on an ancillary space $\cal E$, see corollary~\ref{cor:instrpovm}. If $M(a,A)$ is covariant, then the POVM $P(A)$ is covariant with respect to the representation $D_g$ of $SU(d)$ on $\cal E$,
\begin{equation}
P(g^{-1}A)=D_gQ(A)D_g^*.
\end{equation}

In this section the family of covariant POVMs on the Hilbert space ${\cal E}=\C\otimes \C$ is classified. Holevo's theorem for POVMs $Q(f)\equiv M(\mathbb{I},f)=V^*\mathbb{I}\otimes P(f)V$ yields operators 
\begin{equation}
Q_0=V^*\mathbb{I}\otimes P_0V
\end{equation}
that form the family of POVMs originating from general instruments. At the end of the section the special case $d=2$ is discussed in detail.

\subsubsection{Holevo's Theorem On $\C\otimes\C$}
The family of covariant POVMs is defined by theorem~\ref{the:holprac}.
\begin{theorem}[Holevo's Theorem For Instruments]
\label{the:holprac}
Let $M:\B\otimes \Sigma(\Omega)$ be a covariant instrument and define $Q(A)\equiv M(\mathbb{I},A)$. By the Stinespring dilation theorem and Holevo's theorem, all covariant POVMs $Q(A)$ are defined by operators $Q_0=V^*\mathbb{I}\otimes P_0V$ in which $P_0$ is of the form
\begin{equation}
\label{eq:p0deco}
P_0=\left(\begin{array}{c c}
1 & c \\
\bar{c} & b
\end{array}\right)\oplus eP_{e_1\otimes\mathbb{C}^{d-1}}\oplus fP_{\mathbb{C}^{d-1}\otimes e_1} \oplus gP_{\mathrm{Ad}(d-1)}.
\end{equation}
Normalization yields in addition
\begin{eqnarray}
\label{eq:p0con}
a=1\\
\frac{b}{d^2-1}+\frac{e+f}{d+1}+\frac{gd(d-2)}{d^2-1}=1\\
|c|^2\leq ab.
\end{eqnarray}
\end{theorem}

\emph{Proof:} The operators $P_0$ commute with the stability group $\{D_g;g\in H_{x_0}\}$. The representation space of the stability group of a vector in $\C\otimes\C$ is a subspace of $\bar U\otimes U$. It is given by
\begin{equation}
U(H)\simeq(\bar U(1)\oplus \bar U(d-1))\otimes (U(1)\oplus U(d-1)).
\end{equation}
Operators from this space leave the vector state $\ket{e_1}$ invariant. The representation space of $U(H)$ is 
\begin{align}
(\mathbb{C}e_1\oplus\mathbb{C}^{d-1})\otimes(\mathbb{C}e_1\oplus\mathbb{C}^{d-1})&\simeq\nonumber\\
(\mathbb{C}e_1\otimes \mathbb{C}e_1)\oplus(e_1\otimes\mathbb{C}^{d-1})&\oplus(\mathbb{C}^{d-1}\otimes e_1)\oplus (\mathbb{C}^{d-1}\otimes\mathbb{C}^{d-1}),
\end{align}
and the representation on this space is
\begin{align}
&\mathrm{Triv}(1)\oplus \bar{U}(d-1)\oplus U(d-1)\oplus(\bar{U}(d-1)\otimes U(d-1))\nonumber\\
&\simeq\mathrm{Triv}(1)\oplus \bar{U}(d-1)\oplus U(d-1)\oplus(\mathrm{Triv}(d-1)\oplus\mathrm{Ad}(d-1))\nonumber\\
&\simeq(\mathrm{Triv}\otimes \mathbb{C}^2)\oplus \bar{U}(d-1)\oplus U(d-1)\oplus \mathrm{Ad}(d-1).
\end{align}
In this expression, $\mathrm{Triv}(1)$ and $\mathrm{Triv}(d-1)$ are trivial representations and $\mathrm{Ad}(d-1)$ is the adjoint representation on $\mathbb{C}^{d-1}$. 

The operator $P_0$ commutes with elements of this space, so
\begin{equation}
P_0=M_0\oplus eP_{e_1\otimes\mathbb{C}^{d-1}}\oplus fP_{\mathbb{C}^{d-1}\otimes e_1} \oplus gP_{\mathrm{Ad}(d-1)}
\end{equation}
in which $M_0$ is a Hermitean $2\times 2$~matrix
\begin{equation}
M_0=\left(\begin{array}{c c}
a & c \\
\bar{c} & b
\end{array}\right)
\end{equation}
and the operators $P_{e_1\otimes\mathbb{C}^{d-1}}$,$P_{\mathbb{C}^{d-1}\otimes e_i}$ and $P_{\mathrm{Ad}(d-1)}$ are projections.

Because $P_0$ has to be a positive operator, the coefficients satisfy
\begin{itemize}
	\item $a,b,e,f,g\in\mathbb{R}\qquad a,b,e,f,g\ge 0$
	\item $c\in\mathbb{C}\qquad |c|^2\leq ab.$
\end{itemize}

The variables $a,b,c,e,f,g$ are evaluated with use of the normalization condition eq.~(\ref{eq:p0}). See appendix~\ref{sec:p0norm}. The final result is given by
\begin{eqnarray}
a=1\\
\frac{b}{d^2-1}+\frac{e+f}{d+1}+\frac{gd(d-2)}{d^2-1}=1\\
|c|^2\leq ab.
\end{eqnarray}\begin{flushright}$\Box$\end{flushright}

\subsubsection{Covariant Measurement For $d=2$}
In the qubit case, $d=2$, the action of $H$ leaves the vector state $\ket 0$ invariant and so acts on $\mathbb{C}^2\otimes\mathbb{C}^2$ as $(\bar U(1)\oplus \bar U(1))\otimes (U(1)\oplus U(1))$:
\begin{align}
\left(\begin{array}{c c}
e^{-i\phi} & 0 \\
0 & e^{-i\psi}
\end{array}\right)\otimes
\left(\begin{array}{c c}
e^{i\phi} & 0 \\
0 & e^{i\psi}
\end{array}\right)= 
\left(\begin{array}{c c c c}
1 & 0 & 0 & 0\\
0 & e^{-i(\phi-\psi)} & 0 & 0\\
0&0&e^{i(\phi-\psi)}&0\\
0&0&0&1
\end{array}\right)
\end{align}
The operator $P_0$ has to commute with this representation, i.e.
\begin{equation}
P_0=M_0\oplus eP_{e_1\otimes\mathbb{C}}\oplus fP_{\mathbb{C}\otimes e_1}= 
\left(\begin{array}{c c c c}
a & 0 & 0 & c\\
0 & e & 0 & 0\\
0&0&f&0\\
\bar{c}&0&0&b
\end{array}\right)
\end{equation}
in which $M_0$ is a positive Hermitean $2\times 2$-matrix. Notice that this operator is written in the basis of $\{\psi_0,\psi_0^\perp\}$ given by,
\begin{align}
\psi_0= \frac{1}{\sqrt{2}}\left(\begin{array}{c}1\\0\\0\\1\end{array}\right),
\psi_0^\perp= \frac{1}{\sqrt{2}}\left(\begin{array}{c}-1\\0\\0\\1\end{array}\right).
\end{align}  
The coefficients are calculated with use of the normalization condition, see eq.~(\ref{eq:p0}), 
\begin{eqnarray}
a=1\\
b+e+f=3,\qquad 0 \leq b,e,f\\
|c|^2\leq b.
\end{eqnarray}
\subsubsection{Calculating The Seed}
The operator $Q_0=V^*\mathrm{I}\otimes P_0V$ is calculated explicitly. The Stinespring operator $V$ of a covariant CP is
\begin{equation}
V=\frac{1}{\sqrt{2}}\left(\begin{array}{c c}
\sqrt{1-\alpha} -\frac{\sqrt{a}}{\sqrt{3}}& 0 \\
 0 & 2\frac{\sqrt{a}}{\sqrt{3}}\\
 0 & 0\\
 \sqrt{1-\alpha} +\frac{\sqrt{a}}{\sqrt{3}} & 0 \\
 0 & \sqrt{1-\alpha} +\frac{\sqrt{a}}{\sqrt{3}}\\
 0&0\\
 -2\frac{\sqrt{a}}{\sqrt{3}}&0\\
  0&\sqrt{1-\alpha} -\frac{\sqrt{a}}{\sqrt{3}}
\end{array}\right)
\end{equation}
and $P_0$ in the standard basis is
\begin{equation}
P_0=\frac{1}{2}\left(\begin{array}{c c c c}
 1- 2c + b & 0 & 0 & 1-b\\
 0 & 0&0&0\\
 0 & 0&0&0\\
 1-b & 0&0&1+ 2c + b
\end{array}\right).
\end{equation}
This implies that the seed $Q_0$ is given by
\begin{align}
Q_0&=V^*\mathbb{I}\otimes P_0 V\nonumber\\
&=\left(\begin{array}{c c}
1-\alpha + 2c\frac{\sqrt{\alpha-\alpha^2}}{\sqrt{3}}+\alpha\frac{b+2f}{3} & 0 \\
0&1-\alpha - 2c\frac{\sqrt{\alpha-\alpha^2}}{\sqrt{3}}+\alpha\frac{b+2e}{3}
\end{array}\right).
\end{align}
See chapter~\ref{chap:trade} for details. 

\begin{example}[Rotating Polarizer Measurement]
\label{ex:rotpol2}
Consider the measurement instrument of example~\ref{ex:rotpol}. This measurement device measures the polarization of a photon in a randomly chosen direction. Let's extend this instrument to a device that measures the polarization state in a randomly chosen direction with probability $\lambda$ and does nothing at all with probability $1-\lambda$. The POVM is given by the seed
\begin{equation}
Q_0=\left(\begin{array}{c c}
1+\lambda & 0 \\
0&1-\lambda
\end{array}\right)=(1+\lambda)q + (1-\lambda)q^\perp,\qquad 0\leq \lambda \leq 1,
\end{equation}
with $q=\ket 0\bra 0$ and $q^\perp=\ket 1\bra 1$. Observe that 
\begin{align}
Q(\D p)&=uQ_0u^*\D u\nonumber\\
&=(1+\lambda)uqu^*\D u + (1-\lambda)uq^\perp u^*\D u\nonumber\\
&=(1+\lambda)uqu^*\D u + (1-\lambda)u(\mathbb{I}-q)u^*\D u\nonumber\\
&=(2\lambda)p\D p + (1-\lambda)\mathbb{I}
\end{align}
which implies
\begin{equation}
\label{eq:rotgain}
Q(f)=2\lambda\int_\PO f(p)p\D p + (1-\lambda)\int_\PO f(p)\D p \mathbb{I}.
\end{equation}
The choice $\lambda=1$ clearly corresponds to the POVM of the prior example. The through-going channel of the rotating polarizer measurement device depends on $\lambda$ as well. The instrument that implements this device is given by
\begin{equation}
M(a,f)=2\lambda\int_\PO f(p)pap\D p + (1-\lambda)\left(\int_\PO f(p)\D p\right) a.
\end{equation}
The measurement outcome is obviously obtained by $Q(f)\equiv M(\mathbb{I},f)$. The through-going channel $T(a)\equiv M(a,1)$ is given by
\begin{equation}
T(a)=2\lambda\int_\PO pap\D p + (1-\lambda) a.
\end{equation}
By calculation of $\mathrm{Tr}(\hat\rho T(\hat\rho))$ with use of appendix~\ref{sec:traceint}, it's covariant form is found
\begin{equation}
\label{eq:rotdist}
T(a) = \left(1 - \frac{2\lambda}{3}\right)a + \frac{\lambda}{3}\mathrm{Tr}(a)\mathbb{I}.
\end{equation}

As a preview of chapter~3, examine the trade-off between information gain, given by eq.~(\ref{eq:rotgain}), and distortion loss, given by eq.~(\ref{eq:rotdist}). With help of a proper state reconstruction operation, the amount of information obtained by the POVM can be compared on an equal footing with loss of quality due to distortion. Using the fidelity $F_\mathrm{wc}$ as a figure of merit, the distortion loss is given by
\begin{align}
F_\mathrm{wc}(T)&=\mathrm{inf}_{\hat \rho~\mathrm{pure}~}F(\hat\rho,T(\hat\rho))\nonumber\\
&=\mathrm{Tr}\left(\hat\rho\left(\left(1 - \frac{2\lambda}{3}\right)\hat\rho + \frac{\lambda}{3}\mathrm{Tr}(\hat\rho)\mathbb{I}\right)\right)\nonumber\\
&=\left(1 - \frac{\lambda}{3}\right),
\end{align}
in which the infimum may be ommited because of covariance. If $\lambda$ increases, the fidelity $F_\mathrm{wc}(T)$ decreases and so the distortion increases.

Let $E$ be the estimation operation, constructed by composition of the POVM $Q$ and a reconstruction operation $R$. It turns out that the information gain, so the fidelity $F_\mathrm{E}$, is linear in $\lambda$ as well. Fig.~(\ref{fig:rot}) is an illustration of information gain vs distortion loss of the rotating polarizer measurement apparatus. See chapter~\ref{chap:trade} for more details. 

\begin{figure}[!htb]
\begin{center}
\includegraphics[width=8cm]{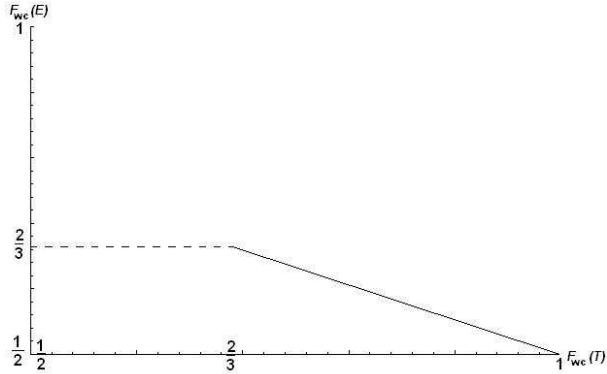}
\caption{Information gain vs distortion loss of measurement in a randomly chosen basis of a $2$-dimensional quantum system. No distortion, i.e. $F_\mathrm{wc}(T)=1$, implies no information gain, $F_\mathrm{wc}(E)=1/2$. The point $F_\mathrm{wc}(T),F_\mathrm{wc}(E)=(2/3,2/3)$ correpsonds to optimal measurement. Note that total distortion of the initial quantum state, i.e. $F_\mathrm{wc}(T)=1/2$, is not realized by this instrument.}
\label{fig:rot}
\end{center}
\end{figure}
\end{example}
\newpage

\chapter{Trade-off}
\label{chap:trade}
This chapter centered around theorem~\ref{the:main} in section~\ref{sec:trade}.
\section{Introduction}

The Heisenberg principle asserts that information extraction from a quantum system is accompanied by distortion losses. It implies that the knowledge a classical observer can acquire about any physical property of a quantum system is limited. Although the many deep implications do not make the Heisenberg principle a founding principle of quantum mechanics, it certainly is a leitmotiv.

Some direct implications of the Heisenberg principle in particular have been studied extensively in the last two decades: the impossibility to estimate the state of $N$ identically prepared quantum systems perfectly and the prohibition of perfect quantum cloning~\cite{Woot}. A quantum system cannot be copied perfectly, since if it could, the state could be fully estimated using statistical measurement on the copies~\cite{Bru}. Upper bounds to optimal cloning have been derived and practical implementations saturating this bound realized~\cite{Exp,Wer1}. 

Many upper bounds to information gain have been derived. The state of an arbitrary quantum system cannot be estimated with 100\% reliability. The mean fidelity $\tilde F$ of a guessed state provided by any estimation scheme is restricted to $\tilde F \leq 2/(d+1)$ with $d$ the dimension of the system. Explicit schemes have been constructed and it turns out that a finite set of measurement operators suffice to saturize this bound~\cite{Der}. 

It is worth noting that the Heisenberg principle is correctly formulated as: ``there exists at least one state, such that, if a system is measured, i.e. information is extracted, and this information is disregarded, this state will be changed.'' Not all states are changed, so in consideration of single quantum systems, for example qubits, the notion of mean fidelity is less valuable. A more appropriate figure of merit should take heed of the ``worst case performance'' of a quantum operation. In this sense, optimal estimation schemes are based on covariant measurement~\cite{Hol}.

When discussing optimal estimation schemes, the distortion of the initial state is mostly disregarded. Although this is the main importance of the Heisenberg principle, the trade-off between information gain and distortion loss of a quantum system has been derived only recently by Banaszek~\cite{Ban}. He derived this upper bound analytically by classification of all Krauss operators defining measurement. 

In this section I will prove the same trade-off independently of the methods applied by Banaszek. The key of the proof is classification of measurement instruments, i.e. quantum operations with two outputs, namely the classical measurement outcome and a conditional state. A central role is played by the family of covariant instruments. An important side result is the classification of this family. Examples of covariant instruments and quantum operations, are the optimal cloning device~\cite{Wer1} and the optimal spin-flipping device~\cite{Gis}.

\newpage
\section{Optimal Trade-Off}
\label{sec:trade}
Consider an instrument $M$ that measures a pure quantum system of finite dimension $d$ in an arbitrary (unknown) state with corresponding density matrix $q$. The output of the instrument is given by a measurement result and a conditional state. The instrument $M$ maps $\Sigma(\Omega) \times M_d$ to $M_d$ which is the set of all $d\times d$-matrices and $\Sigma(\Omega)$ are subsets of the possible outcomes $\Omega$. The measurement result of the instrument is obtained by disregarding the conditional state and is given by the POVM $Q(A): \Sigma(\Omega)\to M_d : A \mapsto M(A,\mathbb{I})$. By disregarding the measurement result we get a completely positive (CP) map, $T:M_d\to M_d: a \mapsto M(\Omega, a)$. In fact $T(\rho)$ corresponds to the averaged state after measurement of a state $\rho$.

In order to judge the quality of the classical of the instrument, we want to treat the measurement result on an equal footing with through-going channel of the instrument. To do so, we will make use of a (hypothetical) reconstruction operation $R$ that reconstructs a pure quantum state according to the measurement result. It reconstructs a pure state, because any mixed state could be trivially constructed by combination of pure states. It is clear that we can choose our set of outcomes $\Omega$ to be the projective space $\PH$, the set of pure states, because labeling the estimation of the state with pure states can be done equally well before and after reconstruction. The POVM $Q$ and the reconstruction operation $R$ together yield the estimation operation $E: M_d\to M_d : a \mapsto Q\circ R(a)= \int_{\PH}\mathrm{Tr}(pa)Q(\D p)$ in which the measure $Q(\D p)$ is defined by $Q(A)=\int_A Q(\D p)$ with $A\subseteq \PH$ and $\D p$ the Haar measure on $\PH$.

As stressed in the introduction, we want to judge the quality of the instrument by evaluating its worst case performance, i.e. its performance in case that the input state is the quantum state that by the Heisenberg principle is maximally distorted. Thus, the figure of merit is the fidelity $F_\mathrm{wc}(O)$ of some operation $O:M_d\to M_d$ and is defined by $F_\mathrm{wc}(O) = \mathrm{inf}_q\mathrm{Tr}(q O(q))$. The infimum is restricted to pure states $q$, since joint-concavity of fidelity implies that mixed states are equally or less distorted by a quantum operation than pure states. Let $\rho=\sum_i \lambda_i \ket i \bra i$ be the initial, mixed state of the system. Then,
\begin{align}
F(\rho,O(\rho))&=F\left(\sum_i \lambda_i \ket i \bra i,O(\sum_i \lambda_i \ket i \bra i)\right)\nonumber\\
&\geq \sum_i \lambda_i F(\ket i \bra i,O(\ket i\bra i)),
\end{align}
and thus there exist at least one $\ket i$ such that
\begin{equation}
F(\rho,O(\rho))\geq F(\ket i,O(\ket i)).
\end{equation}
This implies that $F(\rho,O(\rho))\geq F_\mathrm{wc}(O)$.

The objective is finding the joint-restrictions of the pair $F_\mathrm{wc}(T)$ and $F_\mathrm{wc}(E)$, given by
\begin{eqnarray}
F_\mathrm{wc}(T)=\mathrm{inf}_q\mathrm{Tr}(qT(q)),\\
F_\mathrm{wc}(E)=\mathrm{inf}_q\int_{\PH}\mathrm{Tr}(pq)\mathrm{Tr}(qQ(\D p)).
\end{eqnarray}

\subsection{The Main Theorem}
First of all, no distortion implies no information gain, i.e. 
\begin{equation}
F_\mathrm{wc}(T)=1\Rightarrow F_\mathrm{wc}(E)=\frac 1d.
\end{equation}
Note that a fidelity of $\frac {1}{d}$ is equivalent to the overlap between a pure state and the fully mixed state. As an analogy consider a $2$-dimensional classical system of a coin. The best estimation of the system is a random guess, such that the overlap probability of the state of the coin and an estimated state is $\frac 12$.

Furthermore, the fidelity of optimal estimation is derived to be~\cite{Der, Mas}:
\begin{equation}
F_\mathrm{wc}(E)\leq \frac{2}{d+1}.
\end{equation}
If an instrument provides a guess with this fidelity, the fidelity of the through-going channel of the instrument cannot exceed the value $\frac{2}{d+1}$ either. Indeed, if it could, there would still be information left to be extracted and an additional guess could be made over the distorted state, such that the procedure would improve in optimality. And this is not possible by theorem.

The main result of this section is theorem~\ref{the:main}. It applies for all measurement instruments on finite-dimensional quantum systems.

\begin{theorem}
\label{the:main}
Let $M : \Sigma(\PH) \times M_d \to M_d$ be an instrument with associated POVM $Q(A)=M(A,\mathbb{I}_{\C})$ and averaged state after measurement $T(a)=M(\PH,a)$. Let $E(a)=\int_{\PH}\mathrm{Tr}(pa)Q(\D p)$ be the preparation of a quantum state in accordance with the measurement result. Then the possible values of the pair $(F_\mathrm{wc}(T),F_\mathrm{wc}(E))$ in the quadrant $(\frac{1}{d}\leq F_\mathrm{wc}(T),F_\mathrm{wc}(E) \leq 1)$ consist of two regions:
\begin{enumerate}
	\item	{ 
					\begin{equation}
					\frac{1}{d}\leq F_\mathrm{wc}(E),F_\mathrm{wc}(T)\leq \frac{2}{d+1}
					\end{equation}
				}
	
	\item	{	
					\begin{align}
					&\left(dF_\mathrm{wc}(E)-\frac{2d-2}{d}+\frac{d-2}{d}F_\mathrm{wc}(T)\right)^2 +\nonumber\\ 				  
					&\frac{4(d-1)}{d^2}\left(F_\mathrm{wc}(T)-\frac{d+2}{2(d+1)}\right)^2\leq\frac{d-1}{(d+1)^2}\\
					&\frac{2}{d+1}\leq F_\mathrm{wc}(T)\leq 1.
					\end{align}
				}
\end{enumerate}
\end{theorem}

\subsubsection{Dimension $d=2$}
\begin{figure}[!hb]
\begin{center}
\includegraphics[width=8cm]{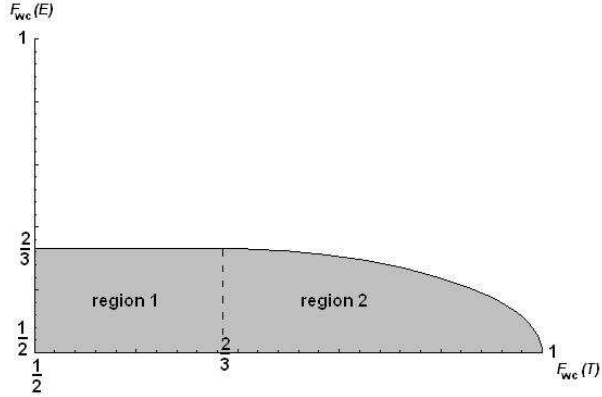}
\caption{The shaded area corresponds to the physically allowed values of $F_\mathrm{wc}(T)$ and $F_\mathrm{wc}(E)$ for dimension $d=2$. The upper bound of region 1 and 2 correspond to optimal measurement. Follow this bound from right to left: the upper bound of region 2 covers information gain and distortion loss of a minimal disturbance measurement (\textsc{MDM}). The upper bound of region 1 corresponds to optimal measurement, ranging from minimal to maximal distortion at the points $(\frac 23, \frac 23)$ respectively $(\frac 13,\frac 23)$. This line expresses the fact that the maximum amount of information extracted out of a quantum system is restricted, regardless of the distortion losses allowed.} 
\label{fig:dim2}
\end{center}
\end{figure}

Fig.~(\ref{fig:dim2}) is an illustration of theorem~\ref{the:main} in case $d=2$.

The pair $(F_\mathrm{wc}(T),F_\mathrm{wc}(E))$ is restricted to two regions:
\begin{enumerate}
	\item	{ 
					\begin{equation}
					\frac{1}{2}\leq F_\mathrm{wc}(E),F_\mathrm{wc}(T)\leq \frac{2}{3}
					\end{equation}
				}
	
	\item	{	
					\begin{align}
					&4\left(F_\mathrm{wc}(E)-\frac{1}{2}\right)^2 + 							
					\left(F_\mathrm{wc}(T)-\frac{2}{3}\right)^2\leq\frac{1}{9},\nonumber\\
					&\frac{2}{3}\leq F_\mathrm{wc}(T)\leq 1.
					\end{align}
				}
\end{enumerate}

The shaded area shows the physical allowed values of $F_\mathrm{wc}(T)$ and $F_\mathrm{wc}(E)$ in the region $1/2\leq F_\mathrm{wc}(T),F_\mathrm{wc}(E)\leq 1$, i.e. the region spanned by all points between full probability overlap of an initial state and the conditional state, respectively estimated state and the probability overlap of an initial state and the fully mixed state. This is the region of physical interest, because $F_\mathrm{wc}(T)\leq 1/2$ corresponds to more distortion than strictly required and $F_\mathrm{wc}(E)\leq 1/2$ can always be attained~\footnote{As an example, let $E'$ be an estimation that produces some state $\hat\sigma$ for any input state. The fidelity $F_\mathrm{wc}$ is a worst case figure of merit. This implies that $F_\mathrm{wc}(E')=0$, while the infimum is reached by the orthogonal state $\hat\sigma\perp$.}.	

\subsection{Proof}

\emph{Proof of theorem~\ref{the:main}:} The key of the proof is classification of all optimal instruments, i.e. all instruments that provide for a fixed value of $F_\mathrm{wc}(T)$ the maximum of $F_\mathrm{wc}(E)$. An important step is the restriction to covariant instruments. Define a rotation operation $\tau_u$ on the estimation operator $E$ by $\tau_u(E)(a)\equiv uE(u^*au)u^*$ in which $u$ is the defining representation of $SU(d)$ on the Hilbert space of the system $\Hi\cong\C$. Let $\tilde{E}$ be defined as the average of $\tau_u(E)$ with respect to the normalized Haar measure $\D u$ on $SU(d)$, i.e. 
\begin{equation}
\tilde{E}\equiv \int_{U(d)} \D u\tau_u (E).
\end{equation}
Note that $F_\mathrm{wc}(E)=F_\mathrm{wc}(\tau_u(E))$ and so because of concavity of $F_\mathrm{wc}$,
\begin{equation}
F_\mathrm{wc}(\tilde{E}) \geq \int_{SU(d)} \D u F_\mathrm{wc} (\tau_u(E)) = F_\mathrm{wc}(E).
\end{equation}
So the average of $\tau_u(E)$ of any estimation operation will provide a guess that is as good as or better than the original non-averaged operation. Similarly for the operator $T$. Thus without loss of generality we can restrict ourselves to $SU(d)$-covariant instruments. 

\subsubsection{Covariant Instruments}
The most general form of an instrument $M(a,A)$ (see section~\ref{sec:covinstr}) is given by its Stinespring dilation:
\begin{equation}
M(a,A)=V^*a\otimes P(A)V,\qquad A\in \Sigma(\Omega),  a\in \B,
\end{equation}
where $P : \Sigma(\Omega) \to {\cal B(E)}$ is a POVM on an ancillary space $\cal E$. 

The averaged output state and the measurement result of an instruments are obtained by defining the quantum operation $T(a)\equiv M(a,\Omega)$ and the POVM $Q(A)\equiv M(\mathbb{I},A)$ such that
\begin{eqnarray}
T(a)=V^*a\otimes \mathbb{I}V,\\
Q(A)=V^*\mathbb{I}\otimes P(A)V.
\end{eqnarray}
If $M(a,A)$ is covariant, then both $T(a)$ and $Q(A)$ are covariant with respect to the action of $SU(d)$. In particular, the POVM $P(A)$ is covariant with respect to a representation $D$ of $SU(d)$ on $\cal E$,
\begin{equation}
P(g^{-1}A)=D_gQ(A)D_g^*.
\end{equation}

The set of covariant through-going channels $T$ is a one-parameter family of quantum operations. See section~\ref{sec:comap}. The through-going channel $T$ is thus given by
\begin{equation}
T(a) = \left(1 - \alpha\frac{d^2}{d^2-1}\right)a + 
\left(\alpha\frac{d}{d^2-1}\mathrm{Tr}(a)\right)\mathbb{I},\qquad 0\leq \alpha \leq 1.
\end{equation}
The strategy in finding restrictions to $(F_\mathrm{wc}(T),F_\mathrm{wc}(E))$ is fixing the former in order to express the latter and maximizing it. 

The fidelity $F_\mathrm{wc}(T)$ is linear in $\alpha$:
\begin{align}
F_\mathrm{wc}(T)&=\mathrm{inf}_q\mathrm{Tr}(qT(q))\nonumber\\
&=\left(1 - \alpha\frac{d^2}{d^2-1}\right)\mathrm{Tr}(q^2) + 
\left(\alpha\frac{d}{d^2-1}\right)\mathrm{Tr}(q)\nonumber\\
&=1-\alpha\frac{d}{d+1}\Rightarrow \alpha=\frac{d+1}{d}\left(1-F_\mathrm{wc}(T)\right).
\end{align}

The estimation fidelity $F_\mathrm{wc}(E)$ will be expressed in $F_\mathrm{wc}(T)$ via a fixed value of $\alpha$. The fidelity of $T$ is to range between $\frac 1d \leq F_\mathrm{wc}(T) \leq 1$, since that are the only physical interesting values. As a consequence, $\alpha$ ranges between \begin{equation}
0 \leq \alpha \leq \frac{d^2-1}{d^2}.
\end{equation}
Larger values of $\alpha$ correspond to state-flipping devices and provide more ``distortion'' than strictly necessary in this setting.

The family of covariant POVMs on the Hilbert space ${\cal E}=\C\otimes \C$ is classified in section~\ref{sec:covinstr}. The POVM $Q(A)=V^*\mathbb{I}\otimes P(A)V$ is via Holevo's theorem (theorem~\ref{the:hol}) given by
\begin{equation}
Q(A)=\int_{gx_o\in A} uQ_0 u^*\D u \equiv\int_A Q(p)\D p
\end{equation}
Herein $Q_0=V^*\mathbb{I}\otimes P_0V$ with $P_0$ given by
\begin{equation}
P_0=\left(\begin{array}{c c}
1 & c \\
\bar{c} & b
\end{array}\right)\oplus eP_{e_1\otimes\mathbb{C}^{d-1}}\oplus fP_{\mathbb{C}^{d-1}\otimes e_1} \oplus gP_{\mathrm{Ad}(d-1)}
\end{equation}
with $0 \leq b,e,f,g$ and
\begin{eqnarray}
\frac{b}{d^2-1}+\frac{e+f}{d+1}+\frac{gd(d-2)}{d^2-1}=1\\
|c|^2\leq ab.
\end{eqnarray}

\subsubsection{Optimization}
The fidelity of the estimation operation is
\begin{equation}
F_\mathrm{wc}(E)=\mathrm{inf}_q\int_{\PH}\mathrm{Tr}(pq)\mathrm{Tr}(qQ(\D p))
\end{equation}
with $Q(\D p)$ by Holevo's theorem given by
\begin{align}
Q(\D p)&=Q(p)\D p=uQ_0u^* \D u\nonumber\\
& = uV^*\mathbb{I}\otimes P_0Vu^* \D u
\end{align}
with
\begin{equation}
Q(uQ_0u^*)\equiv uQ_0u^*.
\end{equation}
It clearly depends on the coefficients $b,c,e,f,g$ of $P_0$ and on $\alpha$ via the Stinespring dilation operator $V$. 

The POVM seed $Q_0$ itself generates a covariant POVM on $\C$. As explained in the text below theorem~\ref{the:hol}, such a POVM is given by
\begin{equation}
Q_0= d\gamma q +\frac{d}{d-1}(1-\gamma)q^\perp.
\end{equation}
Here is $q$ a one-dimensional projection and $\gamma$ some real-valued factor. It follows that
\begin{align}
\mathrm{Tr}(q Q(\D p))&=\braket{e_1}{uQ_0u^* e_1}\D u\nonumber\\
&=\left(d\frac{d\gamma-1}{d-1}\mathrm{Tr}(pq)+\frac{d}{d-1}(1-\gamma)\right)\D p.
\end{align}

The fidelity $F_\mathrm{wc}(E)$ readily depends on $\gamma$ as
\begin{align}
F_\mathrm{wc}(E)&=\mathrm{inf}_q\int_{\PH}\mathrm{Tr}(pq)\mathrm{Tr}(qQ(\D p))\nonumber\\
&=\int_{\PH}d\frac{d\gamma-1}{d-1}\mathrm{Tr}^2(pq)\D p+\int_{\PH}\frac{d}{d-1}(1-\gamma)\mathrm{Tr}(pq)\D p\nonumber\\
&=\frac{\gamma+1}{d+1},
\end{align}
in which we used covariance to omit $\mathrm{inf}_q$ and 
\begin{eqnarray}
d\int_{\PH} \mathrm{Tr}(pq)\mathrm{Tr}(pq)\D p =\frac{2}{d+1}\\
d\int_{\PH} \mathrm{Tr}(pq)\D p =1,
\end{eqnarray}
see appendix~\ref{sec:traceint}. 

The factor $\gamma$ depends, among others, on $\alpha$ and is obtained by calculation of $\gamma=1/d\mathrm{Tr}(qQ_0)$. See appendix~\ref{sec:calgamma}. The result is given by
\begin{align}
\gamma &=\frac{1}{d}\mathrm{Tr}(qQ_0)\nonumber\\
&=\frac{1}{d}\braket{e_1}{Q_0|e_1}\nonumber\\
&=\frac{1}{d}\braket{Ve_1}{\mathbb{I}\otimes P_0|Ve_1}\nonumber\\
&=\frac{1}{d}\mathrm{Tr}_1 (P_{Ve_1}P_0)\nonumber\\
&=\frac{1}{d}\left(1-\alpha + 2c\frac{\sqrt{\alpha-\alpha^2}}{\sqrt{d+1}}+\alpha\frac{b+df}{d+1}\right),
\end{align}
in which $P_{Ve_1}$ is the projection operator $\ket{Ve_1}\bra{Ve_1}$.

Optimization of $F_\mathrm{wc}(E)$ is equivalent to maximization of $\gamma$ over $c,b,e$ for fixed $\alpha$. This is also done in appendix~\ref{sec:calgamma}. The result is given by
\begin{align}
\gamma_\mathrm{max} = \left\{ \begin{array}{ll}
\frac{1}{d}\left(1 + (d-2)\alpha +2\sqrt{d-1}\sqrt{\alpha-\alpha^2}\right) & 0\leq\alpha\leq\frac{d-1}{d}\\
1 & \frac{d-1}{d}<\alpha\leq 1
\end{array} \right.
\end{align}
The theorem is proved by filling in $\gamma_{\mathrm{max}}$ and $\alpha=\frac{d+1}{d}\left(1-F_\mathrm{wc}(T)\right)$. \begin{flushright}$\Box$\end{flushright}

\subsubsection{The Seed Of The Optimal Instrument}
Covariant POVMs are given by a seed $P_0$. The seed of the optimal covariant POVM $Q_0=V^*\mathrm{I}\otimes P_0V$ is calculated explicitly for $d=2$. The Stinespring operator $V$ of sectionn~\ref{sec:comap} is
\begin{equation}
\label{eq:V}
V=\frac{1}{\sqrt{2}}\left(\begin{array}{c c}
\sqrt{1-\alpha} -\frac{\sqrt{\alpha}}{\sqrt{3}}& 0 \\
 0 & 2\frac{\sqrt{\alpha}}{\sqrt{3}}\\
 0 & 0\\
 \sqrt{1-\alpha} +\frac{\sqrt{\alpha}}{\sqrt{3}} & 0 \\
 0 & \sqrt{1-\alpha} +\frac{\sqrt{\alpha}}{\sqrt{3}}\\
 0&0\\
 -2\frac{\sqrt{\alpha}}{\sqrt{3}}&0\\
  0&\sqrt{1-\alpha} -\frac{\sqrt{\alpha}}{\sqrt{3}}
\end{array}\right)
\end{equation}
and $P_0$ in the standard basis is given by
\begin{align}
P_0&=\frac{1}{2}\left(\begin{array}{c c c c}
 1- 2c + b & 0 & 0 & 1-b\\
 0 & 2e&0&0\\
 0 & 0&2f&0\\
 1-b & 0&0&1+ 2c + b
\end{array}\right)\nonumber\\
&=\left(\begin{array}{c c c c}
 2-\sqrt{3} & 0 & 0 & -1\\
 0 & 0&0&0\\
 0 & 0&0&0\\
 -1 & 0&0&2+\sqrt{3}
\end{array}\right).
\end{align}
This implies that the seed $Q_0$ is given by
\begin{align}
Q_0&=V^*\mathbb{I}\otimes P_0 V\nonumber\\
&=\left(\begin{array}{c c}
1+2 \sqrt{\alpha-\alpha^2} & 0 \\
0&1-2\sqrt{\alpha-\alpha^2}
\end{array}\right).
\end{align}
Note that this only applies for $0\leq\alpha\leq\frac 12$. For $\frac 12\leq\alpha\leq $, $Q_0$ is given by
\begin{align}
Q_0 &=\left(\begin{array}{c c}
2 & 0 \\
0&0
\end{array}\right).
\end{align}

In comparison, the seed of the rotating polarizer measurement instrument (see example~\ref{ex:rotpol2}) is given by
\begin{align}
Q_0^\mathrm{pol}
&=\left(\begin{array}{c c}
1+2\alpha & 0 \\
0&1-2\alpha 
\end{array}\right),
\end{align}
with $0\leq\alpha\leq\frac 12$.

\subsection{Lower Bound of Information Gain}
The lower bound of the information gain for a fixed amount of distortion loss in the sense of $F_\mathrm{wc}$ is equal to $0$; an estimation scheme providing an output state independent of the input state will do. That is, the state orthogonal to the output state is the state that yields the lowest fidelity $F_\mathrm{wc}(E)$. Yet this is not a covariant instrument. This imples that the lower bound is physically less relevant. 

However, in the sense of the mean fidelity $\tilde F$, defined as the fidelity averaged over all possible input states, the estimation channel is restricted to a lower bound larger than $0$. Because mean fidelity is rotation invariant and jointly-concave in its input, it is easy to see that the lower bound is saturated by covariant instruments. The lower bound is now found by minimizing
\begin{equation}
\tilde F = \frac{\gamma+1}{d+1}
\end{equation}
over $\gamma$.

\begin{figure}[!hb]
\begin{center}
\includegraphics[width=8cm]{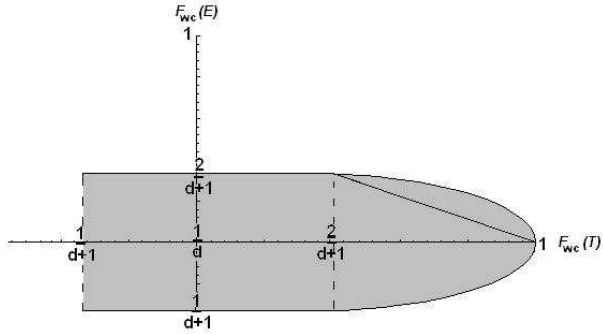}
\caption{The shaded area corresponds to the physically allowed values of $F_\mathrm{wc}(T)$ and $F_\mathrm{wc}(E)$ for covariant instruments on $2$-dimensional systems. The diagonal line corresponds to a rotating polarizer measurement instrument.} 
\label{fig:covariantcigar}
\end{center}
\end{figure}

The calculation of $\gamma_\mathrm{min}$ is similar to the calculation of $\gamma_\mathrm{max}$. The result is given by
\begin{align}
\gamma_\mathrm{min} = \left\{ \begin{array}{ll}
\frac{1}{d}\left(1 +(d-2)\alpha- 2\sqrt{d-1}\sqrt{\alpha-\alpha^2}\right) & 0\leq\alpha\leq\frac{1}{d}\\
0 & \frac{1}{d}<\alpha\leq 1
\end{array} \right.
\end{align}
Fig.~\ref{fig:covariantcigar} illustrates the physical restrictions to the pair $(F_\mathrm{wc}(T),F_\mathrm{wc}(E))$ for all covariant instruments or equivalently of $(\tilde F(T),\tilde F(E))$ for all instruments. Some points in this cigar-like figure are of importance and to be emphasized. The tip of the cigar corresponds of course to complete containment of the initial state, but no information extraction. The points $(\frac 23,\frac 23)$ and $(\frac 23,\frac 13)$ are the fidelities of a device producing an optimal guess respectively an optimal ``anti''-guess of the initial state of the system~\footnote{Since optimal estimation is equivalent to optimal $1\to\infty$-cloning, these devices produce infinitely many clones respectively infinitely many anti-clones of the initial state.}. The boundary at the left corresponds to devices that optimally ``spin''-flip the initial state and in addition yield a measurement result ranging from optimal to ``anti''-optimal. Since an optimal spin-flip device is based on a (classical) measurement scheme~\cite{Wer3}, such a device is equivalent to an instrument that yields an optimal ``anti''-guess, i.e. $F_\mathrm{wc}(E)=\frac 13$. 

The straight diagonal line in fig.~\ref{fig:covariantcigar} is the fidelity trade-off for measurement of the polarization of photons (a qubit-system) in an arbitrary basis. The measurement instrument implementing such measurement consists of a device that measures the polarization state in an arbitrary direction with probability $2\alpha$ and does nothing at all with probability $1-2\alpha$. See for more details examples~\ref{ex:rotpol} and~\ref{ex:rotpol2} in chapter~\ref{chap:qmeas}.

\newpage
\section{Pauli Cloning And Covariant Instruments}
In this section I give a review on the application found by Ref~\cite{Cer}.

\subsection{Introduction}
The article \emph{Separating the Classical and Quantum Information via Quantum Cloning}~\cite{Cer} presents an application of asymmetric quantum cloning. A procedure is constructed to perform a minimal disturbance measurement (\textsc{MDM}) on a 2-dimensional quantum system (qubit). First the qubit is cloned asymmetrically to another system, i.e. a $1\to 2$-cloning device is adopted. Then a generalized measurement is performed on a single clone and an (ancillary) anti-clone or on the two clones. This procedure is used to optimize the transmission of a qubit through a lossy quantum channel. 

It turns out that the optimal measurement instruments as found in chapter~\ref{chap:qm} and in section~\ref{sec:trade} of this chapter is equivalent to the instruments needed for the appliance of the transmission application. 

\subsection{Covariance and Pauli Cloning}
Besides implementing covariant quantum operations, the Stinespring operators $V$ in section~\ref{sec:comap} (see eq.~(\ref{eq:covop})) also implement so called Pauli cloners. The article \emph{Pauli Cloning of a Quantum Bit}~\cite{Cer1} by Cerf introduces this special class of asymmetric cloning machines. Pauli cloners produce two (not necessarily identical) output qubits, each emerging from a Pauli channel. 

\subsubsection{Pauli Channel}
\begin{figure}[!hb]
\begin{center}
\includegraphics[width=8cm]{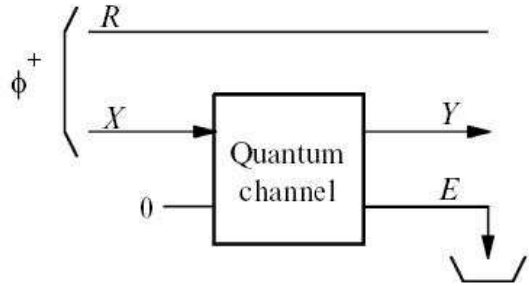}
\caption{A Pauli Channel. The environment $E$ is traced out in order to get the Bell mixture $\rho_{RY}$.}
\label{fig:pauli1}
\end{center}
\end{figure}

Pauli channels act on a qubit in an arbitrary pure state by rotating it by one of the Pauli matrices ($\sigma_x,\sigma_x\sigma_z=-i\sigma_y,\sigma_z$) with probabilities ($p_x,p_y,p_z$) or leaving it unchanged with probability $1-p\equiv 1-p_x-p_y-p_z$. If $p_x=p_y=p_z$, then the Pauli channel is a depolarizing channel. A convenient way of describing the action of a Paul channel is by considering the input qubit $X$ as being maximally entangled with some reference qubit $R$. See fig.~\ref{fig:pauli1}.

Suppose the initial qubit and the reference qubit are initially in the Bell state $\ket{\phi^+}$. The output state of the Pauli channel is then given by the Bell mixture
\begin{equation}
\label{eq:partmix}
\rho_{RY}=(1-p)\ket{\phi^+}\bra{\phi^+} + p_z\ket{\phi^-}\bra{\phi^-}+ p_x\ket{\psi^+}\bra{\psi^+}+p_y\ket{\psi^-}\bra{\psi^-}
\end{equation}
in which $\ket{\phi^-} = 2^{-1/2}(\ket{00}-\ket{11})$, $\ket{\psi^+}=2^{-1/2}(\ket{01}+\ket{10})$ and $\ket{\psi^-}=2^{-1/2}(\ket{01}+\ket{10})$. The fact that $\rho_{RY}$ is this symmetric Bell mixture follows directly from
\begin{eqnarray}
(\mathbb{I}\otimes\sigma_z)\ket{\phi^+}=\ket{\phi^-},\\
(\mathbb{I}\otimes\sigma_x)\ket{\phi^+}=\ket{\psi^+},\\
(\mathbb{I}\otimes\sigma_x\sigma_z)\ket{\phi^+}=\ket{\psi^-}.
\end{eqnarray}
So leaving the reference qubit unchanged and transforming the qubit $X$ with a Pauli operator, yields the state $\rho_{RY}$.

It is clear that a Pauli channel acts on a arbitrary pure state $\rho$ as
\begin{equation}
\rho\mapsto (1-p)\rho + p_x \sigma_x\rho\sigma_x + p_y \sigma_x\sigma_z\rho\sigma_z\sigma_x + p_z\sigma_z\rho\sigma_z
\end{equation}
which for a state-independent Pauli channel, i.e. $p_x=p_y=p_z=p/3$, reduces to
\begin{equation}
\rho\mapsto \left(1 - p\frac{4}{3}\right)\rho + 
\left(p\frac{2}{3}\right)\mathbb{I}\qquad 0\leq p \leq 1.
\end{equation}	
This operation is equal to the general form of covariant quantum operations (see section~\ref{sec:comap}).

\subsubsection{Asymmetric Pauli Cloning}
\begin{figure}[!hb]
\begin{center}
\includegraphics[width=8cm]{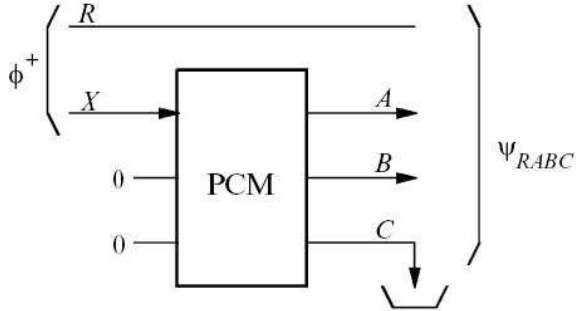}
\caption{A Pauli Cloner. The cloner has two outputs, cloned qubits $A$ and $B$. The ancilla or the environment $C$ is traced out.}
\label{fig:pauli2}
\end{center}
\end{figure}

Pauli cloners are defined as unitary transformations acting on an input qubit $X$ along with two other qubits: the blank copy and an ancillary qubit. Cerf describes the Pauli cloners by considering a 4-qubit system. See fig.~\ref{fig:pauli2}. The initial qubit $X$ is maximally entangled with a reference qubit $R$. Let $X$ and $R$ be in the Bell state $\ket{\phi^+}$. The blank copy and the ancilla are initially in state $\ket 0$. The two outputs $A$ and $B$ admitted by the Pauli cloners are required to emerge from Pauli channels, i.e. the states $\rho_{RA}$ and $\rho_{RB}$ must be Bell mixtures. An ancillary space (the ancilla qubit) is needed by the Schmidt-decomposition. Assume that the Bell state $\rho_{RA}$ results from the partial trace of pure state in an extended space. This extended space needs to be at least 4-dimensional, because it has to accommodate the four eigenvalues of $\rho_{RA}$. The blank copy is thus not sufficient. An extra system of dimension at least $2$ is needed. It is proved by Ref~\cite{Niu} that one ancillary qubit is sufficient to cover optimal asymmetric $1\to\infty$-cloning.

The characterization of the Pauli cloners is based on the requirements that after cloning the states $\rho_{RA}$ and $\rho_{BC}$ are Bell mixtures and that the state of every qubit pair ($RA,RB,RC,AB, AC, BC$) is a Bell smixture as well. This last requirement implies that the ancilla $C$ also emerges from a Pauli channel. The output state of the Pauli cloner is a 4-qubit wave function $\ket{\Phi}_{RABC}$. By the Schmidt-decomposition for the bipartite partition $RA$ vs $BC$ the pure state $\ket{\Phi}_{RABC}$ is given by
\begin{align}
\label{eq:bellmix}
\ket{\Phi}_{RABC}&=\left(\nu+\frac\mu 2\right)\ket{\phi^+}\ket{\phi^+}_{RA;BC} +\frac\mu 2(\ket{\phi^-}\ket{\phi^-}_{RA;BC}\nonumber\\
&+\ket{\psi^+}\ket{\psi^+}_{RA;BC}+\ket{\psi^-}\ket{\psi^-}_{RA;BC}).
\end{align}
These double Bell states for the partition $RA$ vs $BC$ transform into superposition of double bell states for the two other possible partitions ($RB$ vs $AC$ and $RC$ vs $AB$). For example,
\begin{align}
\ket{\phi^+}\ket{\phi^+}_{RB;AC}&=\frac 12 (\ket{\phi^+}\ket{\phi^+}_{RA;BC} + \ket{\phi^-}\ket{\phi^-}_{RA;BC} \nonumber\\
&+ \ket{\psi^+}\ket{\psi^+}_{RA;BC} + \ket{\psi^-}\ket{\psi^-}_{RA;BC}).
\end{align}
It is therefore clear that the operation
\begin{equation}
\ket{\psi^+}\mapsto \nu \ket{\psi^+}\ket{\psi^+}_{RA;BC} + \mu \ket{\psi^+}\ket{\psi^+}_{RB;AC}
\end{equation}
produces the same state as $\ket{\Phi}_{RABC}$ in eq.~(\ref{eq:bellmix}). Moreover, the operator implementing this transformation is equal to the operator $V$ with $\nu=c_1 + \frac{c_2}{\sqrt{3}}$ and $\mu=\frac{-2c_2}{\sqrt{3}}$. Normalization implies $|\mu|^2+\mu\nu+|\mu|^2=1$ or equivalently $|c_1|^2+|c_2|^2=1$. 

The states $\rho_{RA}$ and $\rho_{BC}$ are Bell mixtures as in eq.~(\ref{eq:partmix}), i.e.
\begin{align}
\rho_{RA}&=\left(\nu+\frac\mu 2\right)^2\ket{\phi^+}\bra{\phi^+} + \frac{\mu^2} 4\ket{\phi^-}\bra{\phi^-}+ \frac{\mu^2} 4\ket{\psi^+}\bra{\psi^+}+\frac{\mu^2} 4\ket{\psi^-}\bra{\psi^-}\nonumber\\
\rho_A &= \left(1 - p\frac{4}{3}\right)\rho + \left(p\frac{2}{3}\right)\mathbb{I},
\end{align}
with $\rho$ the original qubit state and $(1-p)=\left(\nu+\frac\mu 2\right)^2=|c_1|^2\equiv 1-\alpha$. The state $\rho_{BC}$ is given by the same expression. The fidelity of the first clone (corresponding to the $A$ qubit) is thus given by
\begin{equation}
F_\mathrm{wc}^A=1-\frac 23 p = 1-\frac 23 \alpha
\end{equation}

As noted above, the double Bell state $\ket{\Phi}_{RABC}$ transforms into superpositions of double Bell states for the two other possible partitions. Table~\ref{tab:bell} contains the amplitudes of $\ket{\Phi}_{RABC}$ in terms of double Bell states for the other partitions.

\begin{table}[!hb]
\caption{Amplitudes of the double Bell states.}
\label{tab:bell}
\begin{center}
\begin{tabular}{|l|c|c|c|c|}
\hline
$\ket{\Phi}_{RABC}$ & $\ket{\phi^+}\ket{\phi^+}$ & $\ket{\phi^-}\ket{\phi^-}$ & $\ket{\psi^+}\ket{\psi^+}$ & $\ket{\psi^-}\ket{\psi^-}$ \\
\hline \hline
$\ket{\Phi}_{RABC}$ & $\left(\nu+\frac\mu 2\right)$ & $\frac \mu 2$ &$\frac \mu 2$ & $\frac \mu 2$ \\
$\ket{\Phi}_{RBAC}$ & $\frac 12 \left(\nu+2\mu \right)$ & $\frac \nu 2$ &$\frac \nu 2$ & $\frac \nu 2$ \\
$\ket{\Phi}_{RCAB}$ & $\frac 12 \left(\nu+\mu \right)$ & $\frac 12 \left(\nu+\mu \right)$ &$\frac 12 \left(\nu+\mu \right)$ & $\frac 12 \left(\nu-\mu \right)$ \\
\hline
\end{tabular}
\end{center}
\end{table}

\begin{figure}[!hbt]
\begin{center}
\includegraphics[width=8cm]{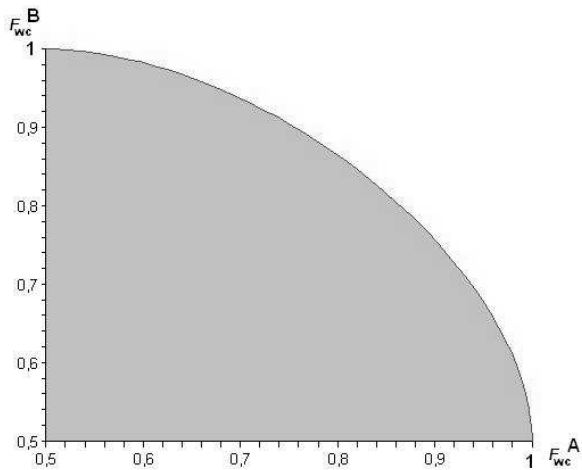}
\caption{The fidelity trade-off between the two clones $A$ and $B$. The plot is restricted to $\frac 12\leq F_\mathrm{wc},F_\mathrm{wc}\leq 1$, because that is the only region of physical interest.}
\label{fig:pauli3}
\end{center}
\end{figure}

This table implies that the state $\rho_{RB}$ (corresponding to the second clone, qubit $B$) is given by
\begin{align}
\rho_{RB}&=\frac 12 \left(\nu+2\mu \right)^2\ket{\phi^+}\bra{\phi^+} + \frac{\nu^2} 4\ket{\phi^-}\bra{\phi^-}+ \frac{\nu^2} 4\ket{\psi^+}\bra{\psi^+}+\frac{\nu^2} 4\ket{\psi^-}\bra{\psi^-}\nonumber\\
\rho_B &=\left(1 - p'\frac{4}{3}\right)\rho + \left(p'\frac{2}{3}\right)\mathbb{I},
\end{align}
with $(1-p')=\frac 14 \left(\nu+2\mu \right)^2=\frac 34 - \frac 12 \alpha \pm \frac 12 \sqrt 3 \sqrt{\alpha-\alpha^2}$. The $\pm$-sign is because $c_2=\pm \sqrt{\alpha}$. This expression is also obtained by explicit calculation of $T'(a)\equiv V^*\mathrm{I}_{\C \otimes\C}\otimes a V$ (see section~\ref{sec:comap}). The fidelity of the second clone is now given by
\begin{equation}
F_\mathrm{wc}^B=1-\frac 23 p' = \frac 12 + \frac 13 \alpha \pm \frac 1{\sqrt 3} \sqrt{\alpha-\alpha^2}.
\end{equation}
Fig.~\ref{fig:pauli3} illustrates the fidelity trade-off between the two clones $A$ and $B$.

The equivalence between the covariant Stinespring dilation theorem and Pauli cloners is as follows. By the Stinespring dilation theorem, every quantum operation is implemented by an isometry which extends the system to a larger space. This is already the essence of asymmetric cloning. Minimal Stinespring dilation for covariant operations yields a class of bounded operators depending on one variable. These operators spread out the initial information over the extended space which is of dimension $2^3=8$. This space is built up of three qubits: the original qubit, which is considered as the first clone, a blank copy, i.e. the second clone and a third qubit, called the ancilla or anti-clone. This ancilla is needed by the Schmidt decomposition. The ``optimal'' spreading out of information (covariant, thus optimal) emerges from Pauli channels, see above. It is equivalent to Pauli cloning defined by Cerf.

\subsection{Application: Separating Classical and Quantum Information}
A direct practical implementation of the trade-off and the instruments saturating optimal trade-off is described by Ref~\cite{Cer}. In this paper the optimal fidelity trade-off is exploited to optimize the transportation of a qubit through a lossy channel. The scheme presented is based on asymmetric Pauli cloning.

The covariant Stinespring operator $V$ actually implements an asymmetric $1\to 2$-cloning device. An initial qubit is asymmetrically cloned into two clones using one ancillary qubit. The fidelity of the first qubit is given by $F_\mathrm{wc}(T)$. The second qubit and the ancilla qubit are optimally measured and provide an optimal guess depending on the quality of the second qubit. The fidelity of this classical information is given by $F_\mathrm{wc}(E)$. The first qubit is sent by a sender, named Arnout, to a receiver, named Bas. In addition Arnout communicates the information he obtained by measurement of the second and ancillary qubit to Bas. If the first qubit does not arrive at Bas', he is compelled to use the classical information which has a fidelity of $F_\mathrm{wc}(E)$. Yet, if it does arrive the fidelity is of course $F_\mathrm{wc}(T)$. 

Let $p$ be the probability that the qubit arrives. The overall fidelity of the transmission is then given by
\begin{equation}
F_\mathrm{cl}=pF_\mathrm{wc}(T) + (1-p)F_\mathrm{wc}(E). 
\end{equation}
Since both the optimal $F_\mathrm{wc}(T)$ and $F_\mathrm{wc}(E)$ depend solely on $\alpha$, this equation can be maximized over $\alpha$ for fixed $p$. See appendix~\ref{sec:toepas}. The result is
\begin{equation}
F_\mathrm{cl}=\frac 16 (3 + p + \sqrt{1+p(5p-2)}).
\end{equation}
This fidelity is optimal since $F_\mathrm{wc}(T)$ and $F_\mathrm{wc}(E)$ saturates the Banaszek bound and the inequalities of theorem~\ref{the:main}. The overall fidelity $F$ is always larger than the fidelity yielded by a device that just sends the qubit with transmittivity $p$ (which yields $F_\mathrm{dir}=\frac{1+p}{2}$).

The transmission reliability could even increase if Arnout is allowed to use quantum memory and Bas can communicate to him whether he received the qubit. If the qubit is lost, Arnout performs an optimal measurement on the second qubit and the ancilla qubit yielding a fidelity of $\frac 23$. If not, Arnout carries out an incomplete Bell measurement, communicates the result to Bas who recovers the original qubit by applying the appropriate Pauli operator. On average, the fidelity is given by $F_\mathrm{qm}=\frac 23 + \frac 13 p$.

\begin{figure}[!hbt]
\begin{center}
\includegraphics[width=8cm]{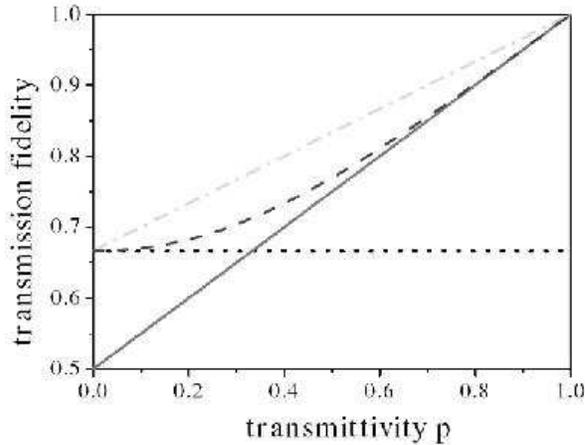}
\caption{The worst case fidelities vs transmittivity. The solid line corresponds to $F_\mathrm{dir}$, the dashed line to $F_\mathrm{dir}$, the dotted line to $F_\mathrm{wc}(E)=\frac 23$ and the dash-dotted line to $F_\mathrm{qm}$.}
\label{fig:pauli4}
\end{center}
\end{figure}

Fig.~\ref{fig:pauli4} illustrates the fidelities of the different transmission strategies. 

The measurement on the second and ancilla qubit is covariant. If the ``worst case'' property of the fidelity is dropped and the figure of merit is the mean fidelity, then the optimal measurement could be implemented by a finite set of measurement operators~\cite{Der}. The cloners needed have been realized by amplification processes and by linear optics. This makes this a physically feasible application.
\newpage

\appendix
\chapter{Appendix}
\section{Rules Of Young Tableaux}
\label{sec:young}
Young tableaux provide some powerful tools for classification of the irreducible representations of $SU(d)$. In this thesis two tools are needed: decomposition of a tensor product of two representations and calculation of the dimension of a representations. Let $d$ be the dimension of the representation space. 

A Young diagram consists of nodes, fitted in rows and columns. The construction of a diagram is constrained to the following rules in order to give a full characterization of the irreducible representations.

\begin{itemize}
	\item The number of nodes in a row must decrease from top to bottom.
	\item The maximal number of rows is $d$. A column with $d$ rows can be omitted from the diagram.
\end{itemize}

A Young tableau is obtained by writing in the $(i,j)$-th box, $d+j-i$, in which $i$ is the row and $j$ is the column. As an example consider the following Young diagram for $d=3$:

\setlength{\unitlength}{0.5cm}
\begin{picture}(4,4)
\put(0.3,0){\line(1,0){2}}
\put(0.3,0){\line(0,1){2}}
\put(0.3,1){\line(1,0){3}}
\put(1.3,0){\line(0,1){2}}
\put(0.3,2){\line(1,0){3}}
\put(2.3,0){\line(0,1){2}}
\put(3.3,1){\line(0,1){1}}
\put(0.7,1.5){\tiny{3}} \put(1.7,1.5){\tiny{4}} \put(2.7,1.5){\tiny{5}}
\put(0.7,0.5){\tiny{2}} \put(1.7,0.5){\tiny{3}}
\end{picture}

\subsubsection{Dimension Of A Young Tableau}
The dimension of a representation $u$ is given by the product of the numbers in the boxes of the corresponding Young tableau divided by the product of the hook lengths. 

The hook-length of a box is the number of boxes right of the box and below the box, plus 1, the box itself. So the hook-lengths of the diagram above, written in the boxes, are:

\setlength{\unitlength}{0.5cm}
\begin{picture}(4,4)
\put(0.3,0){\line(1,0){2}}
\put(0.3,0){\line(0,1){2}}
\put(0.3,1){\line(1,0){3}}
\put(1.3,0){\line(0,1){2}}
\put(0.3,2){\line(1,0){3}}
\put(2.3,0){\line(0,1){2}}
\put(3.3,1){\line(0,1){1}}
\put(0.7,1.5){\tiny{4}} \put(1.7,1.5){\tiny{3}} \put(2.7,1.5){\tiny{1}}
\put(0.7,0.5){\tiny{1}} \put(1.7,0.5){\tiny{1}}
\end{picture}

The dimension of this representation is consequently given by
\begin{equation}
\mathrm{dim}~u=\frac{3\cdot 4\cdot 5\cdot 2\cdot 3}{4\cdot 3\cdot 1\cdot 2\cdot 1}=15.
\end{equation}

The dimension of the adjoint representation on the representation space of dimension $d$ is 
\begin{equation}
\mathrm{dim}~\mathrm{Ad}=\frac{d!\cdot (d+1)}{d \cdot (d-2)!}=(d-1)\cdot(d+1)=d^2-1.
\end{equation}

\subsubsection{Tensor Product Of Two Young Tableaux}

Young tableaux provide us with a second important tool, namely the calculation of the coefficients of the Clebsch-Gordan decomposition of tensor product representations, i.e.
\begin{equation}
u_\alpha\otimes u_\beta=\bigoplus a^\gamma_{\alpha\beta}u_\gamma.
\end{equation}
with $u_{\alpha,\beta,\gamma}$ irreducible representations.

The rules to decompose the tensor product representation of two Young diagrams, are quite simple. See for details Ref~\cite{You}. In this thesis, I need to decompose the tensor representations $\mathrm{Ad}\otimes u$ and $\mathrm{Triv}\otimes u$, with Triv the trivial representation and $u$ the defining representation. The latter is trivial, i.e. $\mathrm{Triv}\otimes u=u$. The former is obtained by legitimately gluing the Young diagram of $u$ to the Young diagram of Ad.

\setlength{\unitlength}{0.5cm}
\begin{picture}(6,4)
\put(0,1){\line(0,1){1}}
\put(0,1){\line(1,0){1}}
\put(0,2){\line(1,0){1}}
\put(1,1){\line(0,1){1}}
\put(1.3,1.3){$(u)~~\otimes$}

\put(4.3,0){\line(1,0){1}}
\put(4.3,0){\line(0,1){2.1}}
\put(4.3,1){\line(1,0){1}}
\put(5.3,0){\line(0,1){2.1}}
\put(4.5,0.4){\tiny{d-1}}
\put(4.8,1.3){\circle*{0.1}}
\put(4.8,1.55){\circle*{0.1}}
\put(4.8,1.8){\circle*{0.1}}
\put(4.3,2.1){\line(1,0){2}}
\put(4.3,2.1){\line(0,1){1}}
\put(4.3,3.1){\line(1,0){2}}
\put(5.3,2.1){\line(0,1){1}}
\put(6.3,2.1){\line(0,1){1}}
\put(4.7,2.5){\tiny{1}}
\put(6.6,1.3){(Ad)}
\end{picture}

This gluing is done by adding the single box of $u$ to all possible other boxes of Ad, such that the new diagram is still a Young diagram. In this case, it can be added to the right of the first row, to the right of the second row, and below the last row. So, in diagrams:

\setlength{\unitlength}{0.5cm}
\begin{picture}(6,4)

\put(0,0){\line(1,0){1}}
\put(0,0){\line(0,1){2.1}}
\put(0,1){\line(1,0){1}}
\put(1,0){\line(0,1){2.1}}
\put(0.2,0.4){\tiny{d-1}}
\put(0.3,1.3){\circle*{0.1}}
\put(0.3,1.55){\circle*{0.1}}
\put(0.3,1.8){\circle*{0.1}}
\put(0,2.1){\line(1,0){3}}
\put(0,2.1){\line(0,1){1}}
\put(0,3.1){\line(1,0){3}}
\put(1,2.1){\line(0,1){1}}
\put(2,2.1){\line(0,1){1}}
\put(3,2.1){\line(0,1){1}}
\put(0.4,2.5){\tiny{1}}

\end{picture}
\setlength{\unitlength}{0.5cm}
\begin{picture}(6,5)

\put(0,0){\line(1,0){1}}
\put(0,0){\line(0,1){2.1}}
\put(0,1){\line(1,0){1}}
\put(1,0){\line(0,1){2.1}}
\put(0.2,0.4){\tiny{d-1}}
\put(0.3,1.3){\circle*{0.1}}
\put(0.3,1.55){\circle*{0.1}}
\put(0.3,1.8){\circle*{0.1}}
\put(0,2.1){\line(1,0){2}}
\put(0,2.1){\line(0,1){2}}
\put(0,3.1){\line(1,0){2}}
\put(1,2.1){\line(0,1){2}}
\put(2,2.1){\line(0,1){2}}
\put(0,4.1){\line(1,0){2}}
\put(0.4,2.5){\tiny{2}}
\put(0.4,3.5){\tiny{1}}

\end{picture}
\setlength{\unitlength}{0.5cm}
\begin{picture}(6,4)
\put(0,0){\line(1,0){1}}
\put(0,0){\line(0,1){2.1}}
\put(0,1){\line(1,0){1}}
\put(1,0){\line(0,1){2.1}}
\put(0.4,0.5){\tiny{d}}
\put(0.5,1.3){\circle*{0.1}}
\put(0.5,1.55){\circle*{0.1}}
\put(0.5,1.8){\circle*{0.1}}
\put(0,2.1){\line(1,0){2}}
\put(0,2.1){\line(0,1){1}}
\put(0,3.1){\line(1,0){2}}
\put(1,2.1){\line(0,1){1}}
\put(2,2.1){\line(0,1){1}}
\put(0.4,2.5){\tiny{1}}
\put(2.7,1.3){$=$}
\put(4,1){\line(1,0){1}}
\put(4,1){\line(0,1){1}}
\put(4,2){\line(1,0){1}}
\put(5,1){\line(0,1){1}}
\end{picture}

In the last diagram, the first column is omitted because it contains $d$ boxes. The decomposition of the tensor product of the adjoint representation and the defining representation contains the defining representation again. The defining representation is of dimension $d$ and the other two representations are of dimension $\frac 12 d(d-1)(d+2)$ and $\frac 12 d(d-2)(d+1)$ respectively. Note that the product of the dimensions of the representations $u$ and Ad equals the sum of the dimensions of the representations contained in the decomposition.

\newpage

\section{Holevo's Normalization Conditions}
\label{sec:p0norm}

First the matrix $M_0$ is evaluated. The vectors
\begin{align}
\psi_0 := \frac{\sum_{i=1}^d e_i\otimes e_i}{\sqrt{d}}\\
\psi_0^\perp := \frac{1}{\sqrt{d-1}}\left[\psi_0 - \sqrt{d}e_1\otimes e_1\right]
\end{align}
form an orthonormal basis in $\mathrm{Triv}\otimes \mathbb{C}^2$. Note that $\C\otimes \C$ is spanned by two subspaces on which $\bar{U}(d-1)\otimes U(d-1)=\mathbb{C}\psi_0 \oplus \mathrm{Ad}(d)$ works irreducibly, so that $P_{\psi_0^\perp}$=$\mathbb{I}_{\mathrm{Ad}(d)}$. In this basis $M_0$ is written as
\begin{equation}
M_0 = aP_{\psi_0} + bP_{\psi_0^\perp} + c\braket{\psi_0}{\psi_0^\perp} + \bar{c}\braket{\psi_0^\perp}{\psi_0}
\end{equation}
in which $P_{\psi_0}$ and $P_{\psi_0^\perp}$ are projections on respectively $\psi_0$ and $\psi_0^\perp$. So
\begin{equation}
\int_G uaP_{\psi_0} u^*\mu(dg) = aP_{\psi_0}
\end{equation}
and
\begin{equation}
\int_G ubP_{\psi_0^\perp} u^*\mu(dg) = \frac{b}{d^2-1}\mathbb{I}_{\mathrm{Ad}(d)}.
\end{equation}
Now
\begin{equation}
\int_G uc \braket{\psi_0}{\psi_0^\perp}u^*\mu(dg) = 0,
\end{equation}
because this operator is a non-invertible intertwiner and by Schur's representation lemma equal to 0. Equivalently 
\begin{equation}
\int_G u\bar{c}\braket{\psi_0^\perp}{\psi_0} u^*\mu(dg) = 0.
\end{equation}
The calculation for the projections $P_{e_1\otimes\mathbb{C}^{d-1}}$,$P_{\mathbb{C}^{d-1}\otimes e_i}$ and $P_{\mathrm{Ad}(d-1)}$ yields
\begin{align}
&\int_G ueP_{e_1\otimes\mathbb{C}^{d-1}}\oplus fP_{\mathbb{C}^{d-1}\otimes e_1} \oplus gP_{\mathrm{Ad}(d-1)}u^*\mu(dg)=\nonumber\\ &
\left(e\frac{d-1}{d^2-1}+ f\frac{d-1}{d^2-1}+g\frac{(d-1)^2-1}{d^2-1}\right)\mathbb{I}_{\mathrm{Ad}(d)},
\end{align}
such that in total the normalization condition eq.~(\ref{eq:p0}) is rewritten as
\begin{equation}
aP_{\psi_0}+\left(\frac{b}{d^2-1}+\frac{e+f}{d+1}+\frac{gd(d-2)}{d^2-1}\right)\mathbb{I}_{\mathrm{Ad}(d)}\\
=P_{\psi_0}+P_{\mathrm{Ad}(d)}=\mathbb{I}.
\end{equation}
This yields
\begin{eqnarray}
a=1\\
\frac{b}{d^2-1}+\frac{e+f}{d+1}+\frac{gd(d-2)}{d^2-1}=1\\
|c|^2\leq ab.
\end{eqnarray}
\newpage

\section{Properties Of The Trace}
\label{sec:trprop}
\begin{theorem}
Let $\hat\rho=\ket\psi\bra\psi$ and $\hat\sigma=\ket\phi\bra\phi$ be two pure states. Then
\begin{equation}
\mathrm{Tr}((\hat\rho\hat\sigma)^2)=\mathrm{Tr}^2(\hat\rho\hat\sigma)
\end{equation}
\end{theorem}
\emph{Proof:} 
\begin{align}
\mathrm{Tr}((\hat\rho\hat\sigma)^2)&=\mathrm{Tr}(\ket\psi\braket\psi\phi\braket\phi\psi\braket\psi\phi\bra\phi)\nonumber\\
&=\braket\psi\phi\braket\phi\psi\braket\psi\phi\braket\phi\psi\nonumber\\
&=\mathrm{Tr}(\ket\psi\braket\psi\phi\bra\phi)\mathrm{Tr}(\ket\psi\braket\psi\phi\bra\phi)=\mathrm{Tr}^2(\hat\rho\hat\sigma)
\end{align}\begin{flushright}$\Box$\end{flushright}
\newpage

\section{Integration Over The Trace}
\label{sec:traceint}
The Haar measure on projective Hilbert space is given by
\begin{equation}
dp=\sin^{2d-3} (\theta)\cos \theta d\theta d{\cal S}_{2d-3}
\end{equation}
in which is $0\leq \theta \leq \pi / 2$ and $d{\cal S}_{2d-3}$ the standard integration measure on a (2d-3)-dimensional unit sphere. Now because $\mathrm{Tr}(pq)=\cos^2 (\theta)$ and $V:= \int_{{\cal P}_1}dp$
\begin{align}
& d\int_{{\cal P}_1} \mathrm{Tr}(pq)\mathrm{Tr}(pq)dp = d\int_{{\cal P}_1} \mathrm{Tr}(pq)^2 dp \nonumber\\
& = \frac{d}{V}{\cal S}_{2d-3}\int_0^{\pi / 2} \cos^5(\theta)\sin^{2d-3}(\theta)d\theta \nonumber\\
& = 2d(d-1)\int_0^{\pi / 2}(1-\sin^2(\theta))(1-\sin^2(\theta))\sin^{2d-3}(\theta) \cos \theta d\theta \nonumber \\
& = 2d(d-1)\Bigg\{\bigg[\frac{\sin^{2d-2}(\theta)}{2d-2}\bigg]_0^{\pi / 2} - 2\bigg[\frac{\sin^{2d}(\theta)}{2d}\bigg]_0^{\pi / 2} + \bigg[\frac{\sin^{2d+2}(\theta)}{2d+2}\bigg]_0^{\pi / 2}\Bigg\} \nonumber\\
& = 2d(d-1)\bigg\{\frac{1}{2d-2} - \frac{1}{d} + \frac{1}{2d+2}\bigg\} \nonumber\\
& = \frac{2}{d+1}.
\end{align}
and
\begin{align}
& d\int_{{\cal P}_1} \mathrm{Tr}(pq)dp = d\int_{{\cal P}_1} \mathrm{Tr}(pq) dp \nonumber\\
& = \frac{d}{V}{\cal S}_{2d-3}\int_0^{\pi / 2} \cos^3(\theta)\sin^{2d-3}(\theta)d\theta \nonumber\\
& = 2d(d-1)\int_0^{\pi / 2}(1-\sin^2(\theta))\sin^{2d-3}(\theta) \cos \theta d\theta \nonumber \\
& = 2d(d-1)\Bigg\{\bigg[\frac{\sin^{2d-2}(\theta)}{2d-2}\bigg]_0^{\pi / 2} - \bigg[\frac{\sin^{2d}(\theta)}{2d}\bigg]_0^{\pi / 2}\Bigg\} \nonumber\\
& = 2d(d-1)\bigg\{\frac{1}{2d-2} - \frac{1}{2d}\bigg\} \nonumber\\
& = 1.
\end{align}
In line 3 we used that $V = \frac{\pi^{d-1}}{(d-1)!}$ and ${\cal S}_{2d-3} = \frac{2\pi^{d-1}}{(d-2)!}$.

\newpage
\section{Evaluation Of Gamma}
\label{sec:calgamma}

Calculate $\gamma=\frac 1d \mathrm{Tr}_1 (P_{Ve_1})$.

\begin{align}
d\gamma&=\mathrm{Tr}_1 (P_{Ve_1})\nonumber\\
&=\sum_j\braket{e_j}{Ve_1}\braket{Ve_1}{e_j}\nonumber\\
&=\sum_j\braket{e_j}{c_1 e_1\otimes\psi_0+\frac{c_2}{\sqrt{d^2-1}}(e_1\otimes\psi_0 - d\psi_0\otimes e_1)}\nonumber\\
&\qquad\braket{c_1 e_1\otimes\psi_0+\frac{c_2}{\sqrt{d^2-1}}(e_1 \otimes\psi_0 - d\psi_0\otimes e_1)}{e_j}\nonumber\\
&=\sum_j\braket{e_j}{\left(c_1+\frac{c_2}{\sqrt{d^2-1}}\right)e_1\otimes\psi_0-\frac{c_2 d}{\sqrt{d^2-1}}\psi_0\otimes e_1}\nonumber\\
&\qquad\braket{\left(c_1+ \frac{c_2}{\sqrt{d^2-1}}\right)e_1\otimes\psi_0-\frac{c_2 d}{\sqrt{d^2-1}}\psi_0\otimes e_1}{e_j}\nonumber\\
&=\sum_j\ket{\left(c_1+ \frac{c_2}{\sqrt{d^2-1}}\right)\psi_0\delta_{j1}- \frac{c_2 d}{\sqrt{d^2-1}}\frac{e_j\otimes e_1}{\sqrt{d}}}\nonumber\\
&\qquad\bra{\left(c_1+ \frac{c_2}{\sqrt{d^2-1}}\right)\psi_0\delta_{j1}-\frac{c_2 d}{\sqrt{d^2-1}}\frac{e_j\otimes e_1}{\sqrt{d}}}\nonumber\\
&=\left|c_1+ \frac{c_2}{\sqrt{d^2-1}}\right|^2\ket{\psi_0}\bra{\psi_0}-\left(\bar{c}_1+ \frac{\bar{c}_2}{\sqrt{d^2-1}}\right)c_2\sqrt{\frac{d}{d^2-1}}\ket{\psi_0}\bra{e_1\otimes e_1}-\nonumber\\
&\qquad\left(c_1+ \frac{c_2}{\sqrt{d^2-1}}\right)\bar{c}_2\sqrt{\frac{d}{d^2-1}}\ket{e_1\otimes e_1}\bra{\psi_0}+\sum_j \frac{d|c_2|^2}{d^2-1}\ket{e_j\otimes e_1}\bra{e_j\otimes e_1}.\nonumber
\end{align}
Now use
\begin{equation}
\psi_0^\perp = \frac{\psi_0 - \sqrt{d}e_1\otimes e_1}{\sqrt{d-1}}.
\end{equation}
Then
\begin{align}
&\left|c_1+ \frac{c_2}{\sqrt{d^2-1}}\right|^2\ket{\psi_0}\bra{\psi_0}-\left(\bar{c}_1+ \frac{\bar{c}_2}{\sqrt{d^2-1}}\right)c_2\sqrt{\frac{d}{d^2-1}}\ket{\psi_0}\bra{e_1\otimes e_1}-\nonumber\\
&\qquad\left(c_1+ \frac{c_2}{\sqrt{d^2-1}}\right)\bar{c}_2\sqrt{\frac{d}{d^2-1}}\ket{e_1\otimes e_1}\bra{\psi_0}+\sum_j \frac{d|c_2|^2}{d^2-1}\ket{e_j\otimes e_1}\bra{e_j\otimes e_1}\nonumber\\
&=\ket{\left(c_1+ \frac{c_2}{\sqrt{d^2-1}}\right)\psi_0 - c_2\sqrt{\frac{d}{d^2-1}}e_1\otimes e_1}\bra{\ldots}+\nonumber\\
&\qquad\sum_{j>1}\frac{d|c_2|^2}{d^2-1}\ket{e_j\otimes e_1}\bra{e_j\otimes e_1}\nonumber\\
&=\ket{c_1\psi_0 + \frac{c_2\psi_0^\perp}{\sqrt{d+1}}}\bra{c_1\psi_0 + \frac{c_2\psi_0^\perp}{\sqrt{d+1}}}+ \sum_{j>1}\frac{d|c_2|^2}{d^2-1}\ket{e_j\otimes e_1}\bra{e_j\otimes e_1}.
\end{align}
Continue with $\mathrm{Tr}_1 (P_{Ve_1})$. This equation is splits up in two:
\begin{align}
&\mathrm{Tr}(P_0\ket{c_1\psi_0 + \frac{c_2\psi_0^\perp}{\sqrt{d+1}}}\bra{c_1\psi_0 + \frac{c_2\psi_0^\perp}{\sqrt{d+1}}})\nonumber\\
&=\braket{c_1\psi_0 + \frac{c_2\psi_0^\perp}{\sqrt{d+1}}}{P_0\left(c_1\psi_0 + \frac{c_2\psi_0^\perp}{\sqrt{d+1}}\right)}\nonumber\\
&=\left(\begin{array}{c c} \bar{c}_1 & \frac{\bar{c}_2}{\sqrt{d+1}}\end{array}\right)
\left(\begin{array}{c c}
1 & c \\
\bar{c} & b
\end{array}\right)
\left(\begin{array}{c} c_1 \\ \frac{c_2}{\sqrt{d+1}}\end{array}\right)
\end{align}
in which I used eq.~(\ref{eq:p0deco}) and following equations and $\psi_0,\psi_0^\perp\in \mathbb{C}^2$. The second part
\begin{align}
&\mathrm{Tr}(P_0\sum_{j>1}\frac{d|c_2|^2}{d^2-1}\ket{e_j\otimes e_1}\bra{e_j\otimes e_1})\nonumber\\
&=\frac{f(d-1)d|c_2|^2}{d^2-1}=\frac{fd|c_2|^2}{d+1},
\end{align}
so that in total
\begin{align}
\braket{e_1}{Q_0e_1}&=\frac{fd|c_2|^2}{d+1}+\left(\begin{array}{c c} \bar{c}_1 & \frac{\bar{c}_2}{\sqrt{d+1}}\end{array}\right)
\left(\begin{array}{c c}
1 & c \\
\bar{c} & b
\end{array}\right)
\left(\begin{array}{c} c_1 \\ \frac{c_2}{\sqrt{d+1}}\end{array}\right)\nonumber\\
&=\left(1-\alpha + 2c\frac{\sqrt{\alpha-\alpha^2}}{\sqrt{d+1}}+\alpha\frac{b+df}{d+1}\right).
\end{align}

\subsubsection{Upper Bound}
Optimization of $F_\mathrm{wc}(E)$ is equivalent to maximization of $\gamma$ over $c,b,e$ for fixed $\alpha$. Because $b$ and $f$ are positive numbers related by the normalization conditions of $P_0$, the coefficients are put to $e,g=0$, $f=(d+1)-b/(d-1)$ and $b=c^2$, such that 
\begin{equation}
\gamma=\frac{1}{d}\left(1+(d-1)\alpha + 2c\frac{\sqrt{\alpha-\alpha^2}}{\sqrt{d+1}}-\alpha\frac{c^2}{d^2-1}\right).
\end{equation}
The maximum is obtained by differentiation to $c$ and is given by 
\begin{equation}
c_{\mathrm{max}}=\frac{d^2-1}{\sqrt{d+1}}\frac{\sqrt{\alpha-\alpha^2}}{\alpha}\leq\sqrt{d^2-1}
\end{equation}
in which the upper bound is a consequence of the normalization conditions. Now at $c_\mathrm{max}$ 
\begin{equation}
\gamma=1.
\end{equation}
The upper bound for $c_\mathrm{max}$ is rewritten as
\begin{equation}
\alpha\geq\frac{d-1}{d}
\end{equation}
such that for $\alpha\leq\frac{d-1}{d}$
\begin{equation}
c_\mathrm{max}=\sqrt{d^2-1}.
\end{equation}

\begin{figure}[!htb]
\begin{center}
\includegraphics[width=8cm]{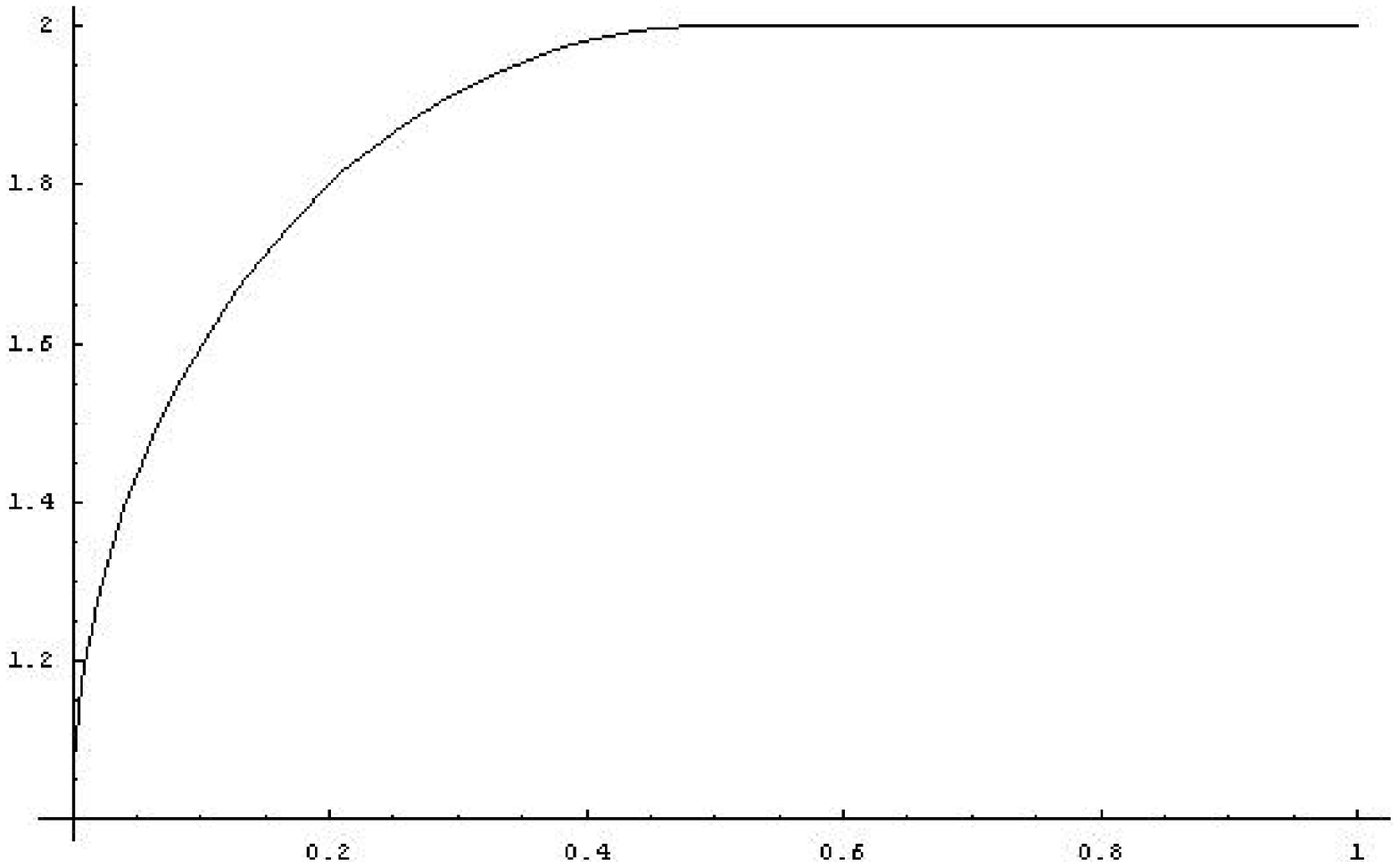}
\caption{Dim=2}
\label{fig:gammadim2}
\end{center}
\end{figure}

The factor $\gamma_\mathrm{max}$ is now given by
\begin{align}
\gamma_\mathrm{max} = \left\{ \begin{array}{ll}
\frac{1}{d}\left(1 + (d-2)\alpha +2\sqrt{d-1}\sqrt{\alpha-\alpha^2}\right) & 0\leq\alpha\leq\frac{d-1}{d}\\
1 & \frac{d-1}{d}<\alpha\leq 1
\end{array} \right.
\end{align}
Fig.~(\ref{fig:gammadim2}) is an illustration of $d\gamma$.

\subsubsection{Lower Bound}
The minimum of $\gamma$ is calculated equivalently. Now put $f=0$ and $b=c^2$. The minimum is reached at
\begin{equation}
c_{\mathrm{min}}=-\frac{d+1}{\sqrt{d+1}}\frac{\sqrt{\alpha-\alpha^2}}{\alpha}\geq -\sqrt{d^2-1}.
\end{equation}
The lower bound for $c_\mathrm{min}$ is rewritten as
\begin{equation}
\alpha\geq\frac{1}{d}
\end{equation}
such that for $\alpha\leq\frac{1}{d}$
\begin{equation}
c_\mathrm{min}=-\sqrt{d^2-1}
\end{equation}
and thus
\begin{align}
\gamma_\mathrm{min} = \left\{ \begin{array}{ll}
\frac{1}{d}\left(1 +(d-2)\alpha- 2\sqrt{d-1}\sqrt{\alpha-\alpha^2}\right) & 0\leq\alpha\leq\frac{1}{d}\\
0 & \frac{1}{d}<\alpha\leq 1
\end{array} \right.
\end{align}
\newpage

\section{Transmission Through A Lossy Quantum Channel}
\label{sec:toepas} 
The maximimum of 
\begin{align}
F_\mathrm{cl}&=pF_\mathrm{wc}(T) + (1-p)F_\mathrm{wc}(E)\nonumber\\
&=p\left(1-\frac 23 \alpha\right) + \frac{1-p}{3}\left(\frac 32 + \sqrt{a-a^2}\right),\qquad 0\leq\alpha\leq\frac 12
\end{align}
is for fixed $p$ reached at
\begin{equation}
\alpha=\frac 12\left(1-\frac{p}{\sqrt{1 + p(5p-2)}}\right).
\end{equation}
This yields
\begin{align}
F_\mathrm{cl}&=pF_\mathrm{wc}(T) + (1-p)F_\mathrm{wc}(E)\nonumber\\
&=\frac 16 (3 + p + \sqrt{1+p(5p-2)}).
\end{align}
\newpage
\newpage

\backmatter


\begin{thebibliography}{99}
\bibitem{Ban} K. Banaszek, Phys. Rev. Lett. 86, 1366, 2001
\bibitem{Bru2} D. Bru\ss, M. Cinchetti, G. M. D'Ariano and C. Macchiavello, Phys. Rev. A 62, 12302, 2000
\bibitem{Bru} D. Bru\ss~and C. Macchiavello, Phys. Lett. A 253, 249, 1999 
\bibitem{Dar} F. Buscemi, G. M. D'Ariano and C. Macchiavello., Phys. Rev. A 71, 2005
\bibitem{Wer3} V. Buzek, M. Hillery, R.F. Werner, Phys.Rev. A 60, 1999
\bibitem{Cav} C.M. Caves, unpublished (available at info.phys.unm.edu/\\ \~{}caves/reports/measures.pdf)
\bibitem{Cer1} N. J. Cerf, Phys. Rev. Lett. 84, 4497, 2000
\bibitem{Dav} E. B. Davies, Quantum Theory of Open System, Academic Press, New York, 1979 
\bibitem{Der} R. Derka, V. Buzek, A. Ekert, Phys. Rev. Lett. 80, 1571, 1998
\bibitem{Emch} G. G. Emch, Mathematical and Conceptual Foundations of the 20th-Century Physics, North-Holland, 1984
\bibitem{Hei} W. Heisenberg,\"Ueber den anschaulichen Inhalt der quantentheoretischen Kinematik and Mechanik, Zeitschrift für Physik 43 172, 1927. English translation in (Wheeler and Zurek, 1983), pp. 62
\bibitem{Gis} N. Gisin and S. Popescu, Phys. Rev. Lett. 83, 432, 1999
\bibitem{Hol} A. S. Holevo, Probabilistic and statistical aspects of quantum theory, Academic Press, 1976
\bibitem{Exp} A. Lamas-Linares et al., Science 296, 712, 2002
\bibitem{Lan} N.P. Landsman,Lecture Notes on C.-Algebras, Hilbert C.-modules and Quantum Mechanics, Amsterdam, 1998
\bibitem{Wer2} M. Keyl and R. F. Werner., J. Math. Phys. 40, 3283, 1999
\bibitem{Kim} G. Kimura, The Bloch Vector for N-level Systems Phys. Lett. A 314, 339, 2003
\bibitem{Man} L. Mandel, Nature 304, 188, 1983
\bibitem{Mas} S. Massar and S. Popescu, Phys. Rev. Lett. 74, 1259, 1995
\bibitem{Mil} P.W. Milonni and M.L. Hardies, Phys. Lett. 92A, 321, 1982
\bibitem{Niels} M. Nielsen and I. Chuang, Quantum Computation and Quantum Information, Cambridge University Press, Cambridge, 2000
\bibitem{Niu} Niu, C.-S., and Griffiths, R. B., 1998, Phys. Rev. A, 58, 4377
\bibitem{Rag} M. Raginsky, J. Math. Phys. 44, 5003-5020, 2003
\bibitem{Cer} M. Ricci, F. Sciarrino, N. J. Cerf, R. Filip, J. Fiuráek, and F. De Martini, Phys. Rev. Lett. 95, 090504, 2005 
\bibitem{Simon} B. Simon, Quantum Dynamics: From Automorphism to Hamiltonian, Studies in Mathematical Physics, Princeton University Press, 1976
\bibitem{Opt} C. Simon, G. Weihs, and A. Zeilinger, Phys. Rev. Lett. 84, 2993, 2000
\bibitem{You} S. Sternberg, Group Theory and Physics, Cambridge University Press, 1994
\bibitem{Tak} M. Takesaki, Theory of operator algebras, I, Springer-Verlag, 1979
\bibitem{Wer} R. F.Werner, Quantum Information Theory - An Invitation (available at the preprint server xxx.lanl.gov,
quant-ph/0101061)
\bibitem{Wer1} R. F. Werner, Phys. Rev. A 58, 1827, 1998
\bibitem{Woot} W. K. Wootters and W. H. Zurek, Nature 299, 802, 1982
\end{thebibliography}
\end{document}